\documentclass[12pt, preprint]{aastex}

\usepackage{amsmath}
\usepackage{amssymb}
\usepackage{graphicx}

\newcommand{\be}{\begin{equation}}
\newcommand{\ee}{\end{equation}}
\newcommand{\bea}{\begin{eqnarray}}
\newcommand{\eea}{\end{eqnarray}}

\begin{document}

\title{The Stellar Number Density Distribution 
in the Local Solar Neighborhood
is North-South Asymmetric}

\author{
Brian Yanny$^{1}$ and 
Susan Gardner$^{2}$ 
}
\affil{
${}^{1}$Fermi National Accelerator Laboratory, Batavia, IL 60510\\
${}^{2}$Department of Physics and Astronomy, 
University of Kentucky, Lexington, KY 40506-0055
}

\begin{abstract}
We study the number density distribution of a sample of K and M dwarf stars, 
matched North and South of the Galactic plane within a 
distance of $2\,{\rm  kpc}$ from the sun, 
using observations from the Ninth Data Release of 
the Sloan Digital Sky Survey. We determine distances using the
photometric parallax method, and in this context 
systematic effects exist 
which could potentially impact 
the determination of the number density profile with height from the 
Galactic plane --- and ultimately affect a number density North-South 
asymmetry. They include: (i) the calibration of the 
various photometric parallax relations, (ii) the ability to separate
dwarfs from giants in our sample, (iii) the role of 
stellar population differences such as age and metallicity, 
(iv) the ability to determine the offset of the sun from the Galactic plane, and (v) 
the correction for reddening from dust
in the Galactic plane, though our 
stars are at high Galactic latitudes. 
We find the various analyzed 
systematic effects to have a negligible impact on our observed asymmetry, and 
using a new and larger 
sample of stars we confirm and refine the earlier discovery of Widrow et al. 
of a significant Galactic North-South asymmetry in 
the stellar number density distribution.  
\end{abstract}



\section{Introduction}

Our own Milky Way Galaxy 
is a spiral galaxy composed of stars, gas, dust --- and
a significant amount of dark matter \citep{gilmore89}. 
While matter beyond that provided by disk stars and gas is 
required to stabilize the Milky Way's disk \citep{op73}, 
the distribution and density 
of that additional, apparently
dark, matter is poorly known (see, e.g., \citet{bt12} and \citet{glrl12}). 
%
Disk systems are not static, and while the literature on the 
azimuthal and radial dynamics of spiral systems is vast, 
less attention has been given to studies of disk dynamics
in the vertical direction.  

The finding of \citet{widrow12} 
(hereafter W12) that there appears to be 
a significant (peak amplitude of $\sim$ 10\%) vertical wave in the number of 
stars within 1 kpc of the sun is surprising and at odds with a history of 
models of the disk as symmetrical and static in the vertical direction. 
It suggests that the Galaxy is not in as close to a state of 
equilibrium in the vertical direction as is frequently assumed, 
and there may be events in the Galaxy's recent past driving this and similar 
vertical waves \citep{getal13}. 
The unexpected nature of this result makes it worth revisiting 
with a larger sample of stars over a larger 
range in Galactic longitude, while paying special attention to possible 
systematic effects which could affect the number counts above and below the 
Galactic plane.

The systematics which we explore are largely connected to our use of
a photometric parallax relation to determine the distance to a star. In this
the intrinsic luminosity of a main-sequence star is inferred from its color, 
so that its distance follows once its apparent magnitude is measured. 
Systematic errors also arise from the confusion of dwarfs with giants, from 
the existence of unresolved binaries, and from stellar population 
differences, particularly in age and metallicity. The stars we consider
are well above the Galactic plane, but we also consider possible inadequacies
in the application of corrections for reddening and 
extinction due to dust. To these ends we 
explore the dereddened colors of distant halo turnoff 
stars as a function of Galactic longitude as well.   
Finally, the location of the sun from the Galactic plane
is also a systematic, in that its value impacts the precise shape of the 
distribution of stars with respect to the Galactic plane, though the inclusion
of this effect does not reduce the significance of the asymmetry we observe. 
Indeed, we find that none of these effects can have 
a significant effect on the large ($\sim$10\%) asymmetry we observe
in the stellar number counts North versus South.


This paper provides the background error analysis for the results of
W12 and precedes the forthcoming paper (Gardner et al., in preparation),
which simultaneously analyzes North-South and left-right \citep{Humphreys11} 
asymmetries and discusses their dynamical implications. 
In anticipation we present here the North-South asymmetry as a function of
Galactic longitude as well. 

\section{Data Selection} 

We employ photometric and astrometric observations from SDSS DR9, the Ninth
Data Release (DR9) of the Sloan Digital Sky Survey (SDSS) \citep{ahn12}.  
DR9 is similar to the Eighth Data Release (DR8) \citep{aihara11}, 
in that it has the same photometric 
imaging footprint, including a significant area of approximately 
1000 square degrees over which the same Galactic longitude range
is covered at moderate Galactic latitudes, avoiding the Galactic plane, 
of equal and opposite sign.  DR9 does correct an earlier issue with the 
astrometry of DR8, which affected positions and proper motions 
of a small subset of stars with $\delta > 41^\circ$. 
In fact, we have used the measured proper motion of M67 to identify
its stars, but that 
cluster is at sufficiently low $\delta$ that it is unaffected
by this change. 
The photometry 
from DR9 is also remeasured separately from earlier SDSS data releases, 
and this has possible effect on the calibration of photometric 
parallax relations. We consider this explicitly. 
One further difference between DR9 and earlier SDSS imaging releases 
is that the DR9 release does not resolve 
the centers of dense clusters into individual stars 
as well as in earlier releases, such as in the Seventh Data 
Release (DR7) \citep{abazajian09}.  
These issues mean that when we test photometric parallax relations on 
known clusters, we find differences between different SDSS data releases.  
These differences, however, are demonstrably small, and in fact have
negligible impact on the asymmetry.

The same telescope, instrument, and software reduction system 
have been used to determine both the North and South data sets, and this 
gives us confidence that most of the systematic effects which could affect 
an asymmetry in the star counts have largely cancelled out. 
Nevertheless, issues remain, 
specifically 
regarding inadequacies in the reddening corrections, which differ in the 
North and South, as well as large-scale 
calibration offset differences, both of which could mimic stellar population differences.
We analyze these in more detail than could be done in the limited
space of W12. 


To minimize obfuscation from dust, we consider high Galactic latitudes, 
so that the viewing is well out of the Galactic plane. We thus 
assume that all of the stars we select are beyond
the dust, which is thought 
to lie at $|z|<125\,{\rm pc}$ \citep{marshall06} --- 
this study is for $l < 100^\circ$ only, but Fig.~19 of 
\citet{berry12} reveals that the dust scale height actually 
increases as one goes to the 
Galactic center, making the assessment of \citet{marshall06} a conservative 
one for our analysis. 
Throughout we convert from the observed apparent magnitudes 
in the SDSS $ugriz$ filters 
to ``dereddened'' ones, denoted by a ``0'' subscript, 
using the extinction maps of \citet{schlegel98}. The dust maps 
depend on the Galactic longitude $l$ and latitude $b$ and are rather
different in the North and South, though the effects of dust are 
substantially reduced at the higher latitudes
which we employ here. Moreover, for our stars, which are
predominately late K and M dwarfs, 
the vast majority of the region in both the north and south has 
extinction $E(B-V) < 0.05$. 
Specifically, using the \citet{schlegel98} maps, we note that 
less than 1\% and 5.6\% of the area in the North and
South, respectively, has $E(B-V) >0.1$. 

\begin{figure}
\centering
\includegraphics[scale=0.45]{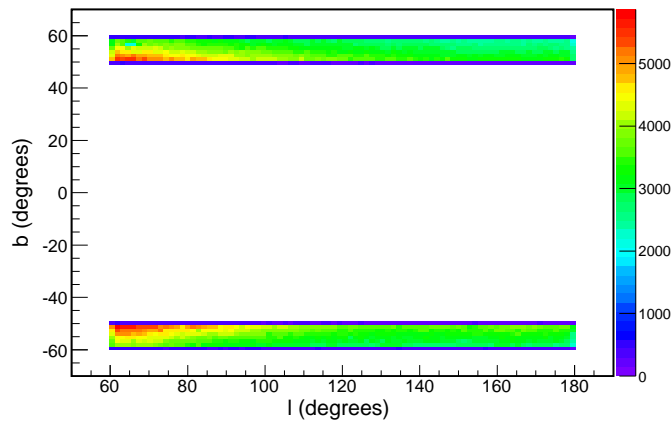}
\caption{
A map of $b$ versus $l$ 
with $13.0 < r_0 < 21.5$ 
for the stars from SDSS DR9 used in this study.
The shading of each bin gives the relative stellar density as indicated by the scale bar to the right of the plot.
}
\label{fig:bvsl}
\end{figure}

We have recently studied (W12) a data set 
selected as per the following criteria: namely, 
stars with Galactic coordinates with 
$100^\circ < l < 160^\circ$ and $54^\circ < |b| <
68^\circ$ that also reside within a perpendicular distance of $1\,{\rm kpc}$
from the line extending directly up and down from the sun in the 
vertical coordinate $z$. 
Limiting the $r_0$-band magnitude such that $14< r_0 < 21$ and 
$0.6 < (r-i)_0 < 1.1$ 
yields a data set of some 300,000 (0.3 M) stars. 
The study of this data set yields a significant North-South 
asymmetry which has the appearance of a wavelike perturbation. 
In this paper we select a similar but non-identical data set, to investigate
whether we can confirm the North-South asymmetry observed in our earlier 
data set. 
Moreover, if an asymmetry is present and is indeed the result 
of a perturbation, it is possible that it is localized 
to particular Galactic longitudes in the neighborhood of the sun.  We will 
pursue this possibility explicitly in Gardner et al. (in preparation), 
and this 
predicates the use of a data set with broader coverage in $l$. 
We thus select the stars of our present study under slightly broader
criteria. The selected stars from SDSS DR9 satisfy the constraints 
$60^\circ < l < 180^\circ$ and $50.3^\circ < |b| < 59^\circ$, with
$13.0 < r_0 < 21.5$ brightness but without saturation 
yielding a sample of some $3.7~\rm M$ stars. 
The selection in $(l,b)$ is shown in Fig.~\ref{fig:bvsl}. 
Restricting to $0.6 < (r-i)_0 < 1.1$ as per W12 yields a data set 
of some $0.85 ~\rm M$ stars, comprised of late K and M dwarfs~\citep{covey07}. 
 We select $l, b$, $r_0$ magnitude, 
$(u-g)_0$, $(r-i)_0$, and $(g-i)_0$ colors for each star.
Note that there is no cut on the distance from the sun imposed explicitly here
as there was in W12. For our asymmetry analysis as a function of Galactic 
longitude, we consider a data set with $0.95 < (g-i)_0 < 2.7$, a sample
of some 1.98M stars, comprised of K and M dwarfs~\citep{covey07}. 

\begin{figure}
\centering
\includegraphics[scale=0.55]{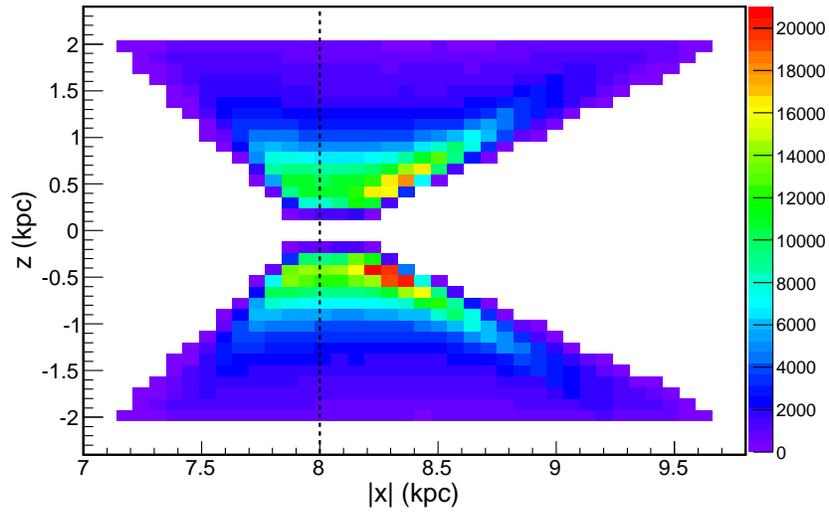}
\caption{
Plot of $z$ versus $\rm |x|$ for the stars from Fig.~\ref{fig:bvsl} in a coordinate system where the Sun is assumed to be located at $\rm |x|= 8$ kpc. 
Our data set samples stars both inside and outside the solar circle. 
}
\label{fig:zvsR}
\end{figure}

The covered region in $z$ and $x$ is shown in Fig.~\ref{fig:zvsR}; 
note that this range is also manifestly North-South symmetric and extends 
to $|z| \sim 2$ kpc.  In comparison, the fits of W12 extend to $|z| =1.6$ kpc. 
We use $r_0$ band observations to assess the apparent magnitude of 
our stars because for the red stars 
we employ this band offers the greatest signal-to-noise ratio for the range of 
types of stars, which are K and M dwarfs, in our sample.
From these inputs we estimate the distance to each star, albeit with 
systematic and statistical uncertainties, as we detail in the next section. 

\section{Photometric Parallax Stellar Distances} 

The color of a main-sequence star can be used to determine its 
absolute magnitude through a photometric parallax relation.  
Although typical random errors for stars at $r=21.5$ in the SDSS catalog 
are 7\%, after averaging together large numbers of measurements we can reduce 
the color errors for stars as faint as this magnitude to the 
level of the systematic error of 1.5\% \citep{Petal08} or below.
Combining statistical and systematic errors in quadrature, 
the associated colors are thus determined to within 3\% 
for even the faintest objects. 

We determine the distance to a star in our sample of stars from DR9 using 
relations constructed using the stars imaged in 
SDSS DR7, such as those 
of \citet{juric08}, in $(r-i)_0$ color, and of \citet{ivezic08}, 
in $(g-i)_0$ color, respectively. 
To employ the photometric parallax method we have made a number of implicit  
assumptions, whose assessment 
ultimately lead to estimates of systematic errors. 
They concern i) the {\it identification} of the stars in the survey, i.e., 
whether they are single, main-sequence stars or not, 
ii) the neglect of {\it unmeasured properties} such as stellar metallicity 
or age on the determined distance, iii) the role of {\it dust} in 
the context of our use of the \citet{schlegel98} maps, and iv) {\it large-scale
calibration} issues between non-contiguously connected regions of the sky. 
We will consider each of these possibilities, and in each case 
we provide detailed, observational tests of our procedures. 

To begin with, however, we test and calibrate the photometric
parallax relations we employ by applying them to the study of 
stellar clusters at known distance.  Of these, we choose clusters out 
of the Galactic plane containing stars of comparable color and 
metallicity to those in our data set.  These criteria severely limit 
the number of clusters we can employ; indeed, the 
clusters  M67 (NGC 2682) and NGC 2420 turn out to be the 
only suitable candidates.  
Both clusters have been imaged (once) by the SDSS and were released
in both DR7 and DR9.  
The image processing was similar but distinct in the two cases. The primary 
difference was that the processing in DR7 
ran with a longer timeout, so that it 
resolved stars appearing deeper towards the centers of these clusters 
than in DR9 \citep{aihara11}. The final photometric 
calibrations of DR7 and DR9 were performed separately using 
methods described in \citet{Petal08}.
We calibrate the photometric parallax relations we use with cluster data 
from DR7. 
Note that the relations used in this paper and in W12 differ
in that different procedures to assess cluster membership were employed
in the two cases. Here we advocate 
a more refined procedure, as we shall describe. 
We also evaluate the impact of the shift from the use of DR7 to DR9 
reductions quantitatively. 

We employ photometric parallax relations devised
by \citet{juric08} in  $(r-i)_0$ color and by \citet{ivezic08}
in $(g-i)_0$ color. 
The analyses of \citet{juric08} and \citet{ivezic08} 
are based on main sequence fitting
to a set of cluster fiducials covering a large range in color. 
In W12 we used a different zeroth coefficient in each case in order to 
confront the literature distance to M67 successfully. 
We have refined that adjustment in our current work. 
For $(r-i)_0$ color, we begin with 
the ``bright'' relation of \citet{juric08}, valid
over the color range $0.1 < (r-i)_0 < 1.6$, include a
correction for the metallicity, and adjust the 
constant term in order to confront the cluster distances
we have noted. In W12 we employed 
\begin{eqnarray}
M_r &=& \Delta M_r ([\hbox{Fe/H}]) + 2.38 + 13.30(r-i)_0 \nonumber \\
&& -11.50(r-i)_0^2+5.40(r-i)_0^3-0.70(r-i)_0^4 \,
\label{pprmi0}
\end{eqnarray}
for the absolute magnitude $M_r$, noting a 
 metallicity correction 
$\Delta M_r ([\hbox{Fe/H}])= -1.11 [\hbox{Fe/H}] - 
0.18 [\hbox{Fe/H}]^2$ \citep{ivezic08},  
with $[\hbox{Fe/H}]$ in dex characterizing the metallicity. 
We assume a universal $[\hbox{Fe/H}] = -0.3$ in our analysis 
of K and M dwarf field stars; however, as we proceed out of the
Galactic plane in $|z|$ we expect the fraction of stars 
(i.e. a thick disk population) with 
a metallicity content of $[\hbox{Fe/H}] \sim  -0.8$ to increase. 
If $[\hbox{Fe/H}]= -0.8$, the assessed distance would be off by some
200 pc, or some 20\% on 1 kpc. Our assessed distance error 
at any given height is a small fraction of this. 
This additional systematic error in distance 
for low-metal, thick-disk stars
could potentially impact the location of the second peak in our 
asymmetry somewhat.
We note, then, that the final 
constant 
term in color for these stars evaluates to 
some 0.50 mag less than that employed in \citet{juric08}. 
Although this may seem like a large offset it will not turn out to be so. 
One reason for this is that the stars used to calibrate \citet{juric08} and
\citet{ivezic08} were much bluer and essentially disjoint from the red stars 
used to calibrate our version
of their photometric parallax relations.  Interested readers can find more sophisticated M-dwarf photometric parallax relations in \citet{bochanski13}.
In performing the current analysis, we realized that the 
large offset we found was driven by our simple assessment of 
cluster membership in M67 and NGC 2420, which we amend here. 
In the current paper we 
employ a constant term of 3.2 in Eq.~(\ref{pprmi0}), 
which is just that of \citet{juric08}, 
though we allow for a metallicity correction of 
$\Delta M_r([\hbox{Fe/H}])$ as well, unlike that work. 
Note that both the current analysis and that of W12 
were adjusted to yield the literature distances
to the clusters; the play of some 20\% in the distance assessments 
is a consequence of the manner in which we deduced cluster membership --- it is removed 
once the membership criteria are sharpened. Thus we investigate 
these issues in careful detail. 
The $(g-i)_0$ relation of \citet{ivezic08}, 
for $0.2 < (g-i)_0 < 4.0$, 
comes 
from confronting DR7 SDSS globular cluster 
data with $(g-i)_0 \lesssim 1$, and the shape 
of the extension to redder colors was developed 
from 
the ``bright'' relation of \citet{juric08}, 
over the color range $0.1 < (r-i)_0 < 1.6$. 
For $(g-i)_0$ color,
we start with the 
relation of \citet{ivezic08} and 
adjust the 
constant term 
in order to confront the cluster distances
we have noted. In W12 we employed 
\begin{eqnarray}
M_r &=& \Delta M_r ([\hbox{Fe/H}]) -1.22 + 14.32(g-i)_0 \nonumber \\
&& -12.97(g-i)_0^2+6.127(g-i)_0^3-1.267(g-i)_0^4+0.0967(g-i)_0^5 \,,
\label{ppgmi0}
\end{eqnarray} 
where we used precisely the same metallicity correction in 
both color bands. We note that our constant 
term for these stars  
is some 0.66 mag less than that employed in \citet{ivezic08}. 
This shift, too, is a consequence of our assessment of cluster
membership. In the current paper we employ a constant of -.50, which is 
0.06 mag greater than the constant in \citet{ivezic08}. 
Finally, we estimate the distance $d$ (in parsecs) to a star with intrinsic 
magnitude $M_r$ and apparent
magnitude $r_0$ via the standard relation $r_0 - M_r = -5 + 5 \log_{10} d$. 
The analysis of \citet{juric08} has been used to study
the stars of the local solar neighborhood, particularly to the end of 
determining the 
height of the sun with respect to the Galactic plane. 
In comparing the stars selected for our asymmetry analysis 
with those of that study, 
our sample is not as red in color so that it does not go as close to 
the Galactic plane. Moreover, our estimated offset of the intrinsic
luminosity, as per the reference 
distance to stars of a more limited color range, 
also pushes our stars somewhat farther away.

In what follows we reexamine these photometric parallax relations in 
detail 
using the observations of main 
sequence K and M dwarf stars in our two chosen reference clusters,
paying careful attention to cluster membership and other systematic effects. 
We calibrate the photometric parallax relations we use with cluster data 
from DR7. We then employ our calibrated relations to study the distance to the 
stars from the same clusters in the DR9 data set, to examine whether any 
additional adjustments of the photometric parallax relations are necessary. 
We wish to realize the sharpest assessment of the photometric distance
possible, making analysis choices driven by data independent of our
asymmetry sample in order to realize an accurate assessment of its shape.

\begin{figure}
\begin{center}
\includegraphics[angle=-90,scale=0.7]{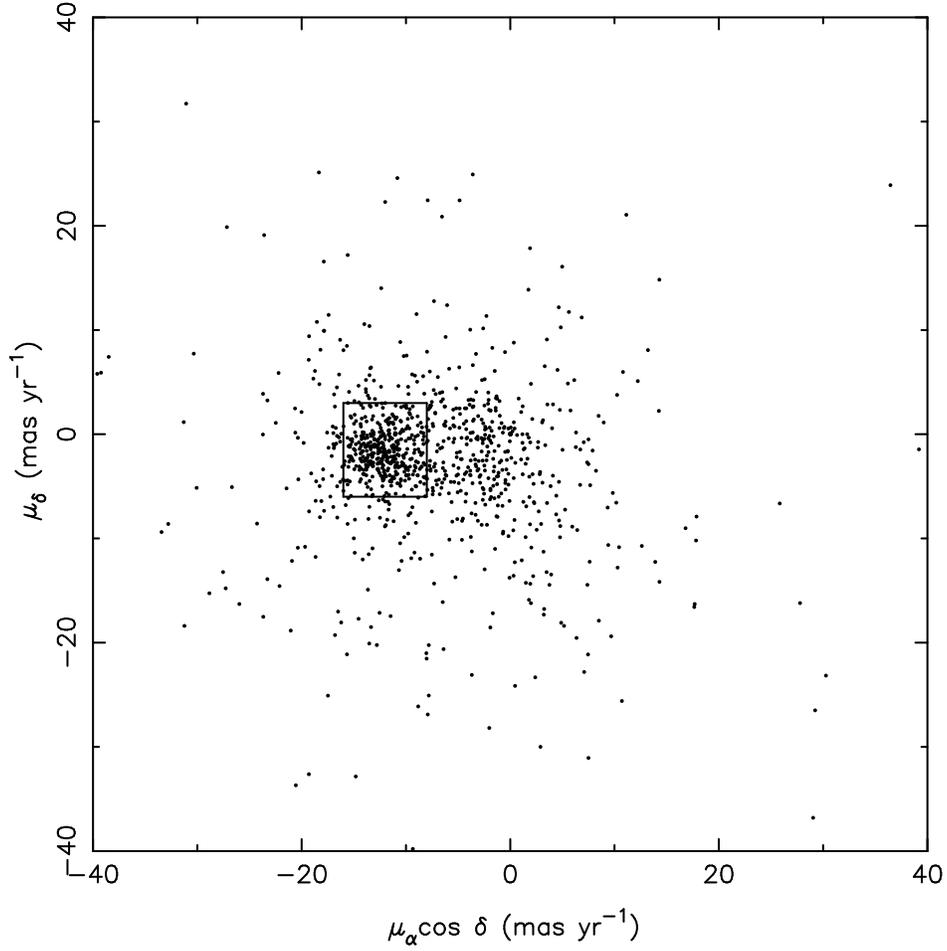}
\caption{
Proper motion of stars in the M67 field.
The box indicates our M67 membership cuts: 
$-16 < \mu_{\alpha} \cos \delta < -8 \> \rm mas\>yr^{-1}$ 
and $-6 < \mu_{\delta} < 3\> \rm mas\>yr^{-1}$. 
Only objects with proper motion errors 
$\sigma(\mu_\alpha \cos\delta,\mu_\delta) < 5 \> \rm mas\>yr^{-1}$ 
are included in the the M67 sample. 
}
\label{fig:m67pm}
\end{center}
\end{figure}

\begin{figure}
\begin{center}
\includegraphics[scale=0.55]{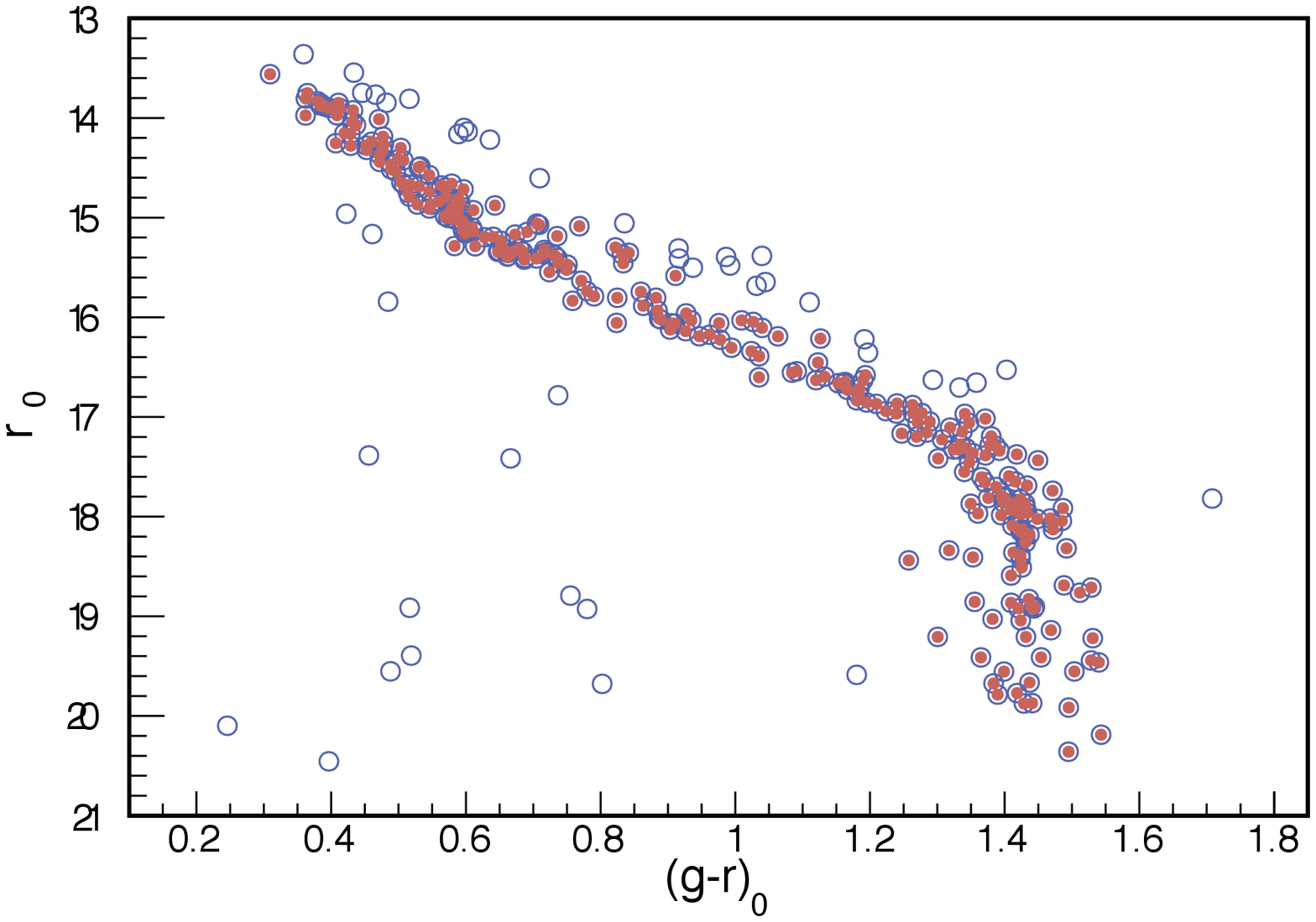}
\caption{
A color magnitude diagram of SDSS DR7 stars presumed to be in M67 
from the proper motion selection box 
in Fig.~\ref{fig:m67pm}.  Filled circles represent objects 
which lie
along the stellar locus and which 
are neither equal-mass binaries nor obvious field stars --- they represent our
cleaned M67 membership sample.
}
\label{fig:hr_m67}
\end{center}
\end{figure}

We begin 
with the cluster M67, for which proper motion information is 
available (and which we did not exploit in W12). 
Figure \ref{fig:m67pm} shows the proper motion of stars in and around the
field of M67.  There are two distinct loci, that of stars in M67 itself 
and that of a somewhat broader
set of stars clearly separated from M67.
These are Milky Way field stars. The preliminary membership cut was based on an
empirical look at the one-dimensional histograms of $\mu$ 
in each coordinate ($\mu_\alpha \cos \delta$ and $\mu_\delta$). 
The cut was slightly refined by displaying candidate members in a 
color-magnitude diagram and finding a reasonably small number 
of outliers. 
We have made the following proper motion cuts: 
$-16 < \mu_{\alpha} \cos\delta < -8\> \rm mas\> yr^{-1}$ and 
$-6 < \mu_{\delta} < 3\> \rm mas\> yr^{-1}$.  
No objects with a proper motion error 
greater than 5 $\rm mas\>yr^{-1}$ in either coordinate are included. 
A histogram of proper-motion errors 
shows a modal value of about 2.5 $\rm mas\>yr^{-1}$.
Figure \ref{fig:hr_m67} shows a ($(g-r)_0,r_0$) color-magnitude 
diagram (CMD) of DR7 
stars near the position of the cluster M67 which 
have proper motions consistent with it. 
A cursory glance at Fig.~\ref{fig:m67pm} shows 
that the Milky Way field in this direction has an average proper 
motion significantly 
offset from the M67 average proper motion, so that 
this procedure removes most field stars, but there is some overlap.

While the proper motion cut significantly decreases outliers, there remain
several field stars which stand out in the CMD.  Also, there is a clear
sequence of unresolved, roughly equal-mass binaries sitting approximately
0.7 mag above the main sequence.  These, as well as the obvious 
field stars, are separated with
simple diagonal cuts and are indicated with open symbols. The filled
circles represent our ``cleaned'' sample used to calibrate our photometric
parallax relations. 
Roughly half of all stars are in binaries, 
as borne out by explicit study of the M67 cluster \citep{fan96}. 
However, since the luminosity of a main-sequence star scales as its mass
raised to the power $3.5$, 
the distance to a sufficiently unequal-mass binary can nevertheless 
be reasonably evaluated through use of a photometric parallax 
relation --- upon assuming that the heavy companion were alone present. 
A star which is 2.5 times brighter than another is approximately 1.0 less 
than it in magnitude; thus 
we see that evaluating the magnitude of a binary of main-sequence stars
by replacing it with the heavier star of them 
yields magnitude shifts of 0.75, 0.57, 0.41, 
and 0.28 for a binary with stars in a mass 
ratio of 1.0, 0.9, 0.8, and 0.7, respectively. 
The appearance of binaries smears the locus of points in the connection of 
color to intrinsic luminosity and thus limits the precision to which we can 
assess the distance --- this will be reflected in the scatter of distances
we determine to the stars in the cluster, which we will assess as a
statistical, or random, error. 
Studies of the mass-ratio distribution shows that binaries 
occur more frequently in relatively close mass pairs, 
both in the M67 cluster, noting Fig.~14 of \citet{fan96}, 
and for solar-like stars in the Milky-Way 
field, noting Fig.~16 of \citet{raghavan10}. 
The presence of such like-mass
binaries smear the color-luminosity relation the most and thus, 
from the perspective of distance assessment, 
are the most problematic; however, from visual inspection of 
Fig.~\ref{fig:hr_m67} we see that they 
are only some 10\% of our cluster stars. 
We identify 27 of such objects in DR7, which we presume are binaries
with mass ratios of 0.9 or greater; this number compares closely
to the some 30 stars noted in Fig.~14 of \citet{fan96}. 
Although equal-mass binaries must also be present in our field stars, 
possibly with equal probability, we have omitted correcting for 
a fraction of these 
objects in our calibration of our photometric parallax relation 
because we believe it best to 
sharpen the distance determination of the preponderance of 
stars in our sample. 
We note that this cleaning procedure was not effected in 
the calibration used in W12, so that we can test 
the impact of this choice through the comparison of our 
results with those of W12. Let us remark in advance 
that we observe no changes which
would be predicated by this. 

%
%
%

\begin{figure}
\begin{center}
\includegraphics[scale=0.55]{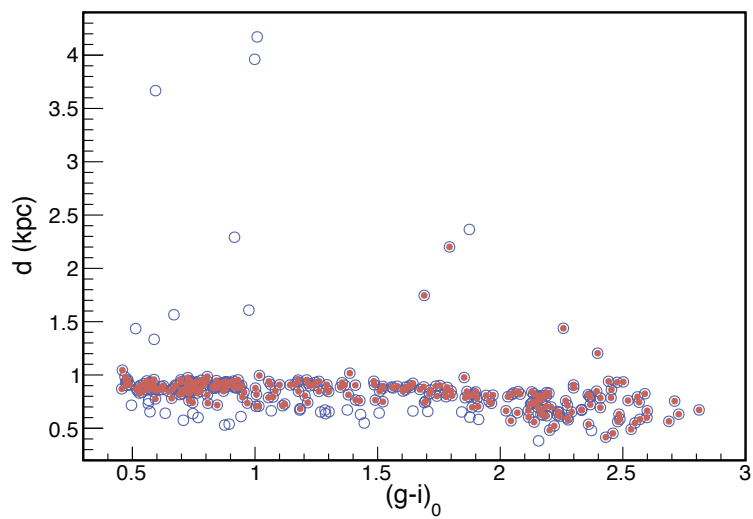}
\caption{
The proper motion selected 
(all circles) and cleaned (filled circles) DR7 samples from
Fig.~\ref{fig:hr_m67}, now plotted with photometric 
parallax distance versus $(g-i)_0$ color. 
There is a slight trend with color for these equidistant points.
}
\label{fig:dgmi_m67}
\end{center}
\end{figure}

\begin{figure}
\begin{center}
\includegraphics[scale=0.55]{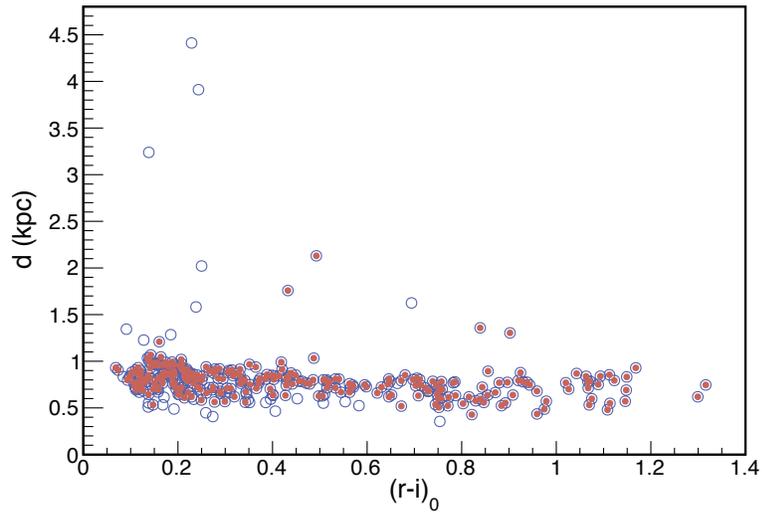}
\caption{
Same as Fig.~\ref{fig:dgmi_m67} but using our 
$(r-i)_0$ based photometric parallax relation.
}
\label{fig:drmi_m67}
\end{center}
\end{figure}

Figure \ref{fig:dgmi_m67} shows a ($(g-i)_0, d$) diagram after applying 
the photometric parallax relation of Eq.~(\ref{ppgmi0}) and employing 
$\rm [Fe/H] = 0.0$ \citep{anmetal08} for the cluster.  
We also show the $((r-i)_0, d)$ 
diagram in Fig.~\ref{fig:drmi_m67}, 
which results from using the distance relation of Eq.~(\ref{pprmi0}). 
As for the reference distance to this cluster, 
we note the recent compilation of \citet{schlesinger12}; assuming
the results therein are independent, we make an error-weighted average,
assuming the errors are independent, 
to find $d_{\rm M67}=0.89 \pm 0.06\,{\rm kpc}$. We use this result 
in what follows, though 
we note by comparison the distance obtained from the use of 
empirically calibrated (theoretical) isochrone models is 
$d=0.84\pm 0.01 $ kpc \citep{anetal07}. 
%

\begin{figure}
\begin{center}
\includegraphics[scale=0.4]{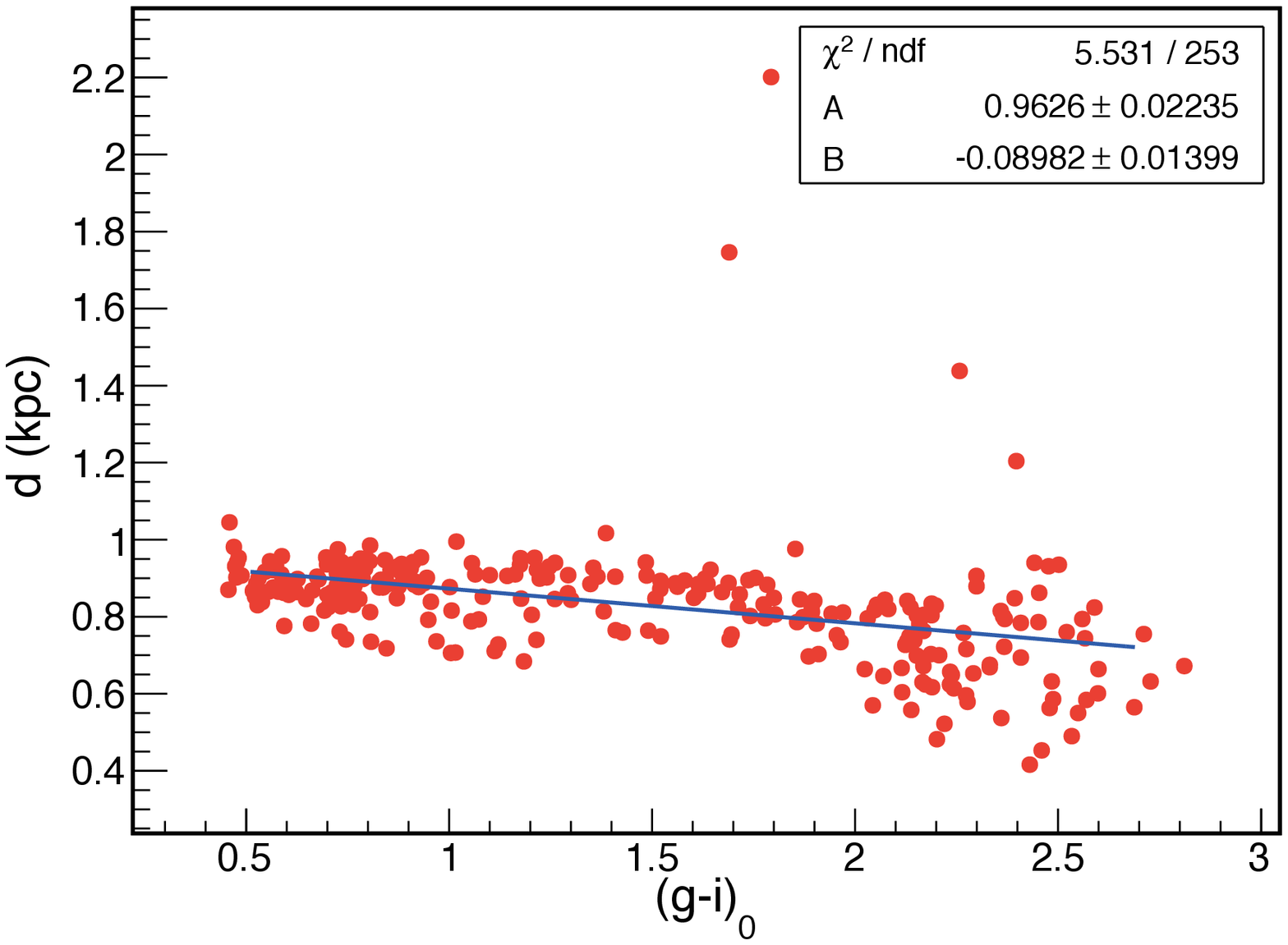} 
\includegraphics[scale=0.4]{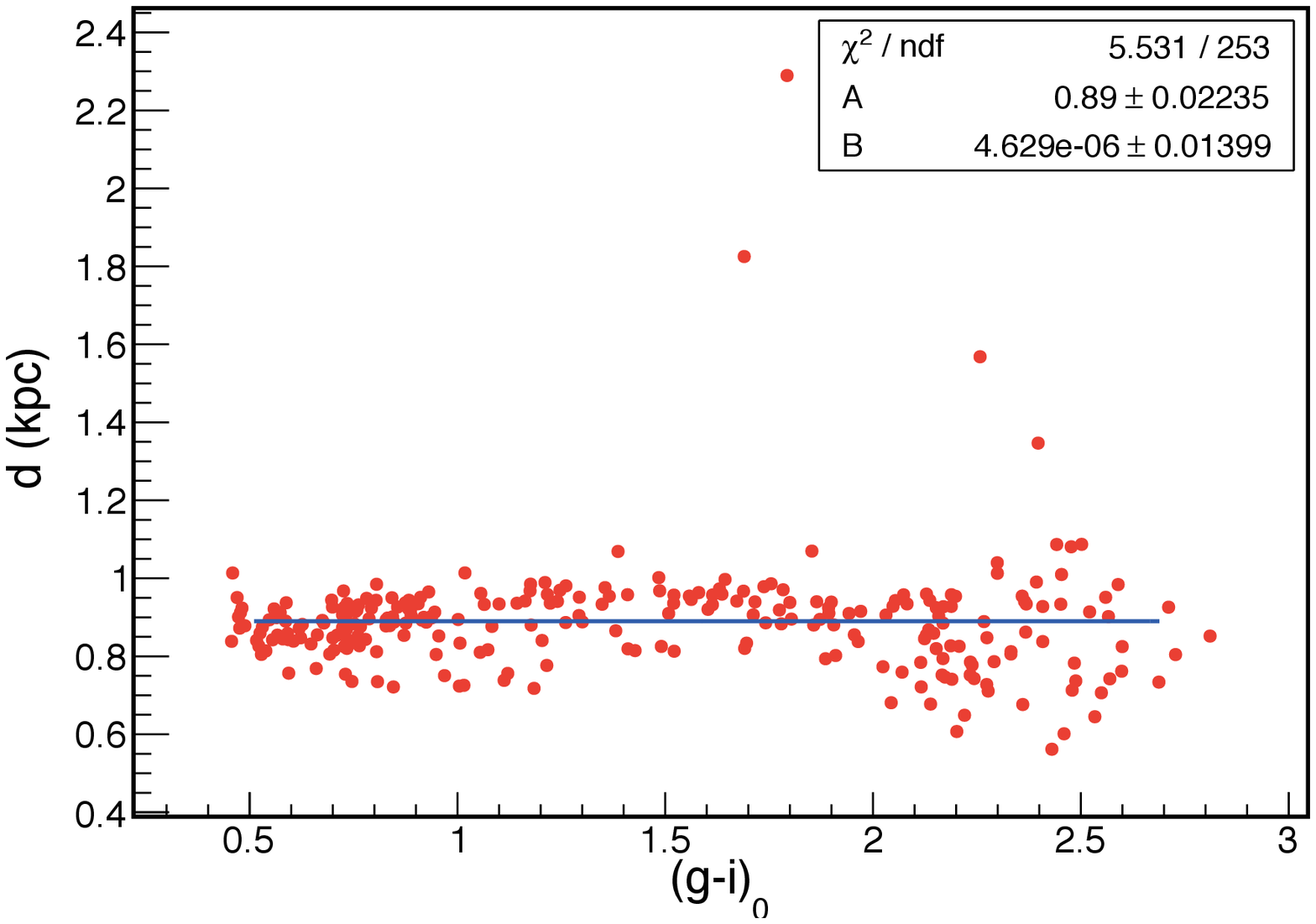}
\includegraphics[scale=0.4]{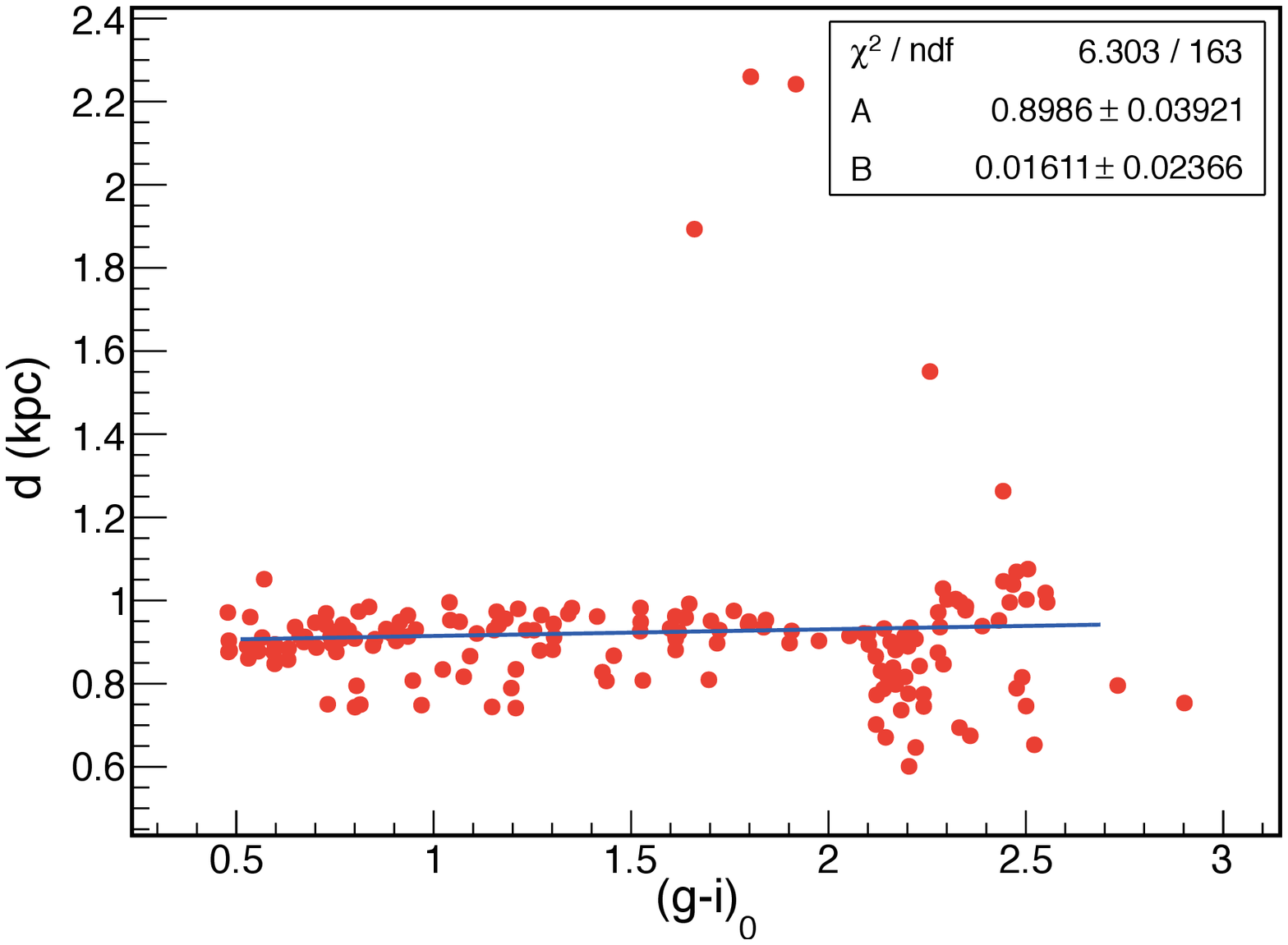}
\caption{
Our final steps in 
refining the $(g-i)_0$ photometric parallax relation 
and assessing systematic and statistical errors:  
the upper left panel shows the fit to the cleaned 
DR7 M67 sample of Fig.~\ref{fig:dgmi_m67}.  After removing a 
linear slope to establish our adopted photometric parallax relation, the 
updated distances are fit and shown in the panel at the upper right.  
The small inset box lists the fit parameters (see text).  
The lower panel shows the updated photometric parallax relation 
determined from the DR7 fit applied to the DR9 sample, after
a similar cleaning procedure has been applied.  A and B refer to the
intercept and slope of the fit respectively.
}
\label{fig:m67gmifix}
\end{center}
\end{figure}

Ideally the determined photometric 
distance to a cluster should be independent
of the employed color; the extent to which this is practically so is a test
of the photometric parallax relation. 
In Fig.~\ref{fig:m67gmifix} we probe the 
$(g-i)_0$ color-distance relationship for M67; 
we show three panels, the first employing Eq.~(\ref{ppgmi0})
to yield $d$ (in kpc), which we
denote as $d[r_0, M_r( (g-i)_0,[{\rm Fe/H}])]$. 
Fitting a linear 
polynomial in $(g-i)_0$ color to these results, for 
$0.5 \le (g-i)_0 \le 2.7$,  yields a 
small correction amounting largely to a slope adjustment. 
Specifically we note the corrected distance $d$ is
\begin{equation}
d 
 = d[r_0, M_r( (g-i)_0,[{\rm Fe/H}])]
+ 0.08982 (g-i)_0 - 0.0726   \,, 
\label{dgmi0tune}
\end{equation}
and the impact of the use of this correction is shown in the second panel. 
Our experience with the $(g-i)_0$ fits 
indicated that the use of a linear correction was sufficient. 
The third panel shows the result of 
applying Eqs.~(\ref{ppgmi0}) and (\ref{dgmi0tune}) to 
the DR9 (rather than the DR7) sample.  
It is apparent that the small correction gives a significantly
better fit, and it is also apparent that the SDSS calibration of the DR9
sample is close enough to the DR7 sample that the same
photometric parallax relations (with linear slope correction) may be 
used for both DR7 and DR9. We now proceed to determine how precisely
we can assess the cluster distance.

\begin{figure}
\begin{center}
\includegraphics[scale=0.33]{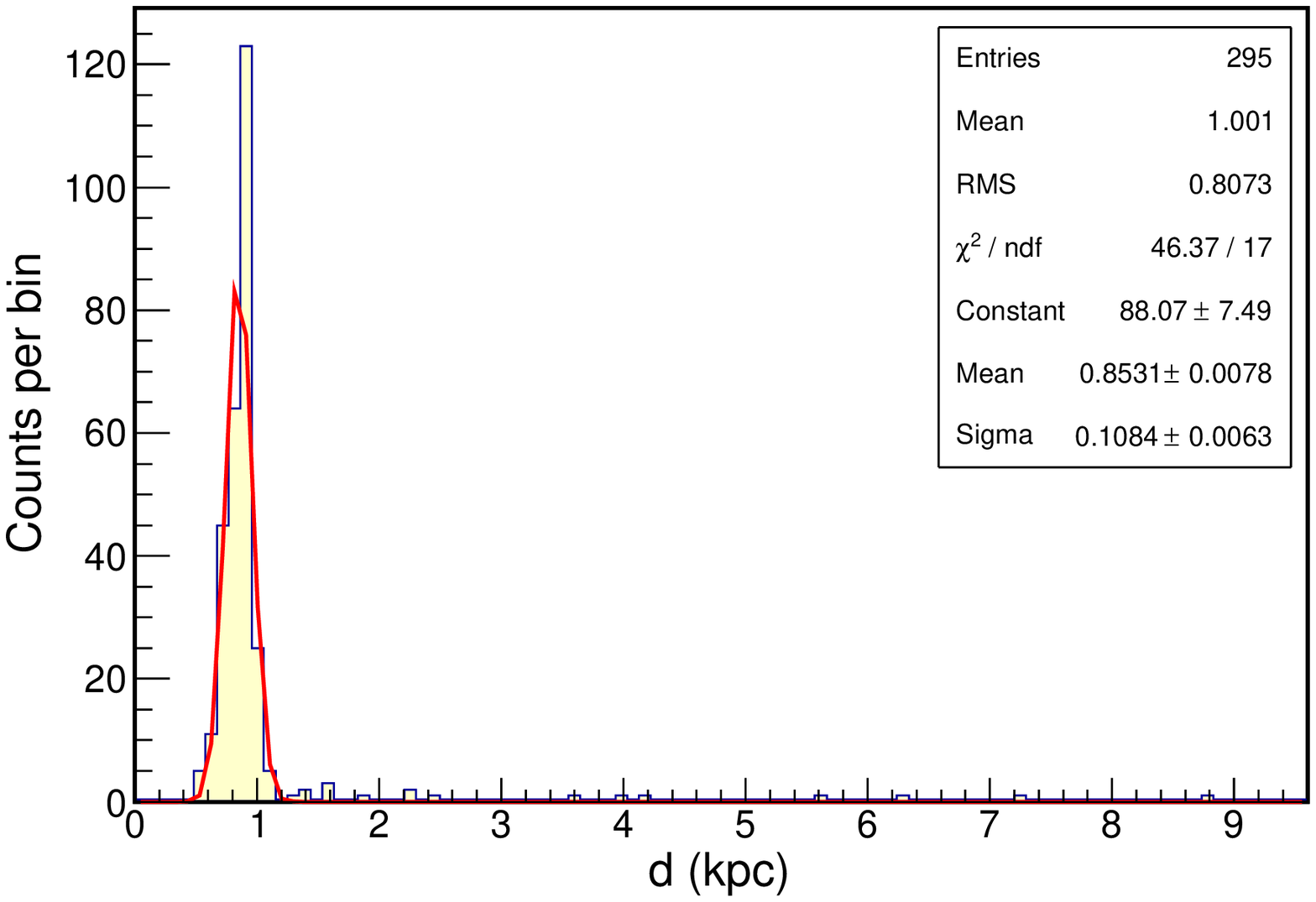}
\includegraphics[scale=0.33]{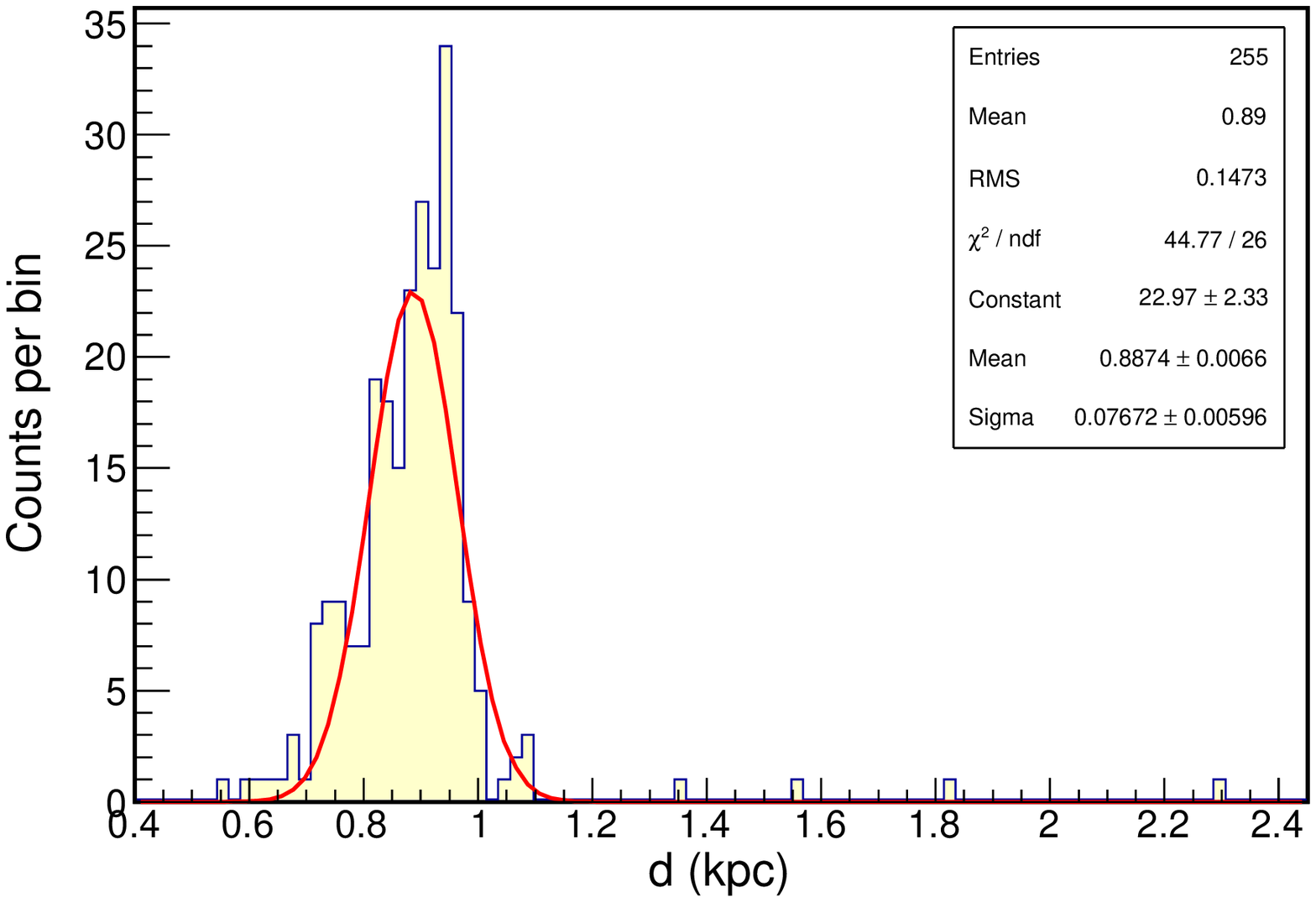}
\includegraphics[scale=0.33]{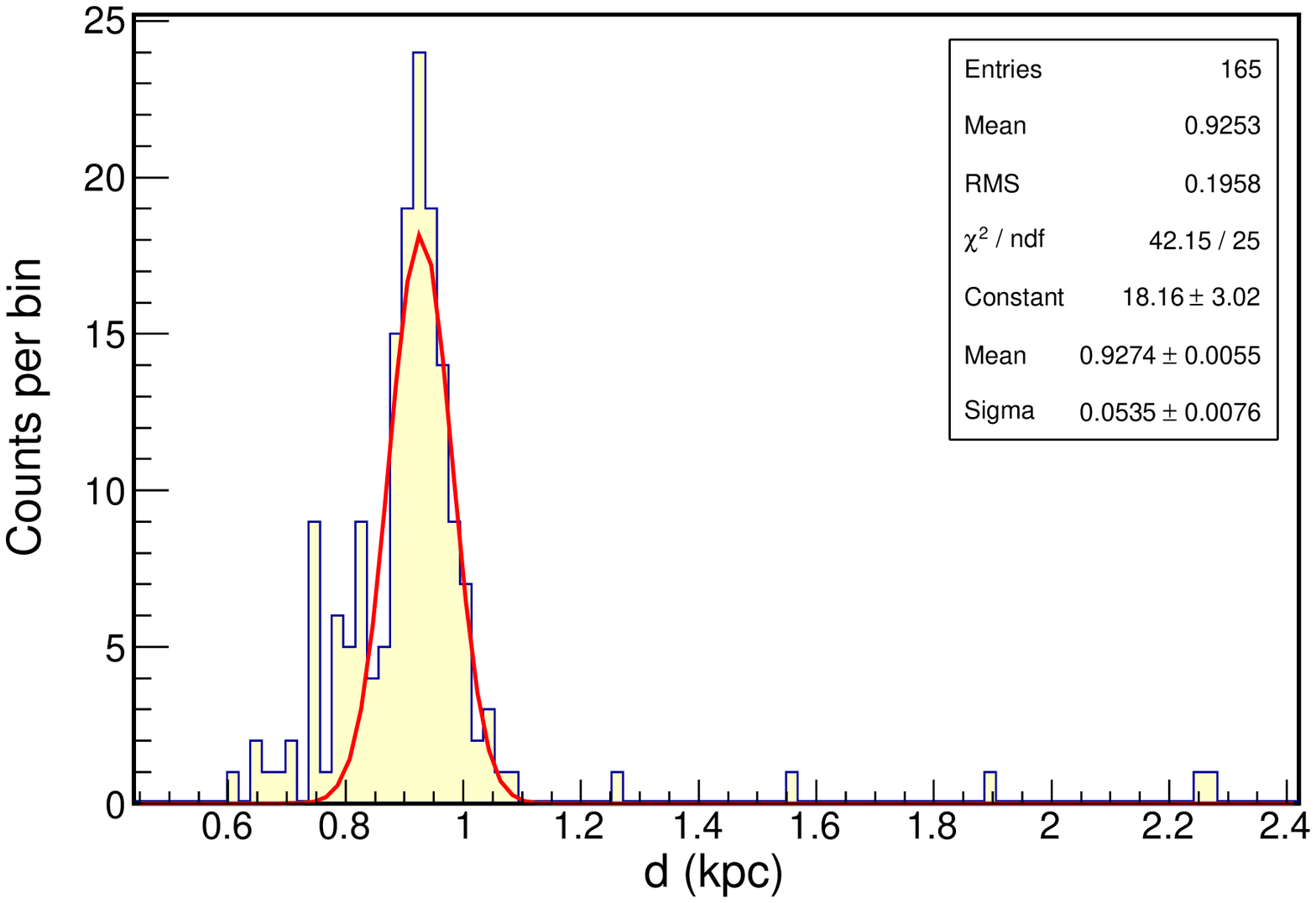}
\caption{
Gaussian fits to  histograms of distance, 
starting with the complete (uncleaned) data set of Fig.~\ref{fig:dgmi_m67}
and using our refinement of Eq.~(\ref{ppgmi0}) for the stellar distances. 
The RMS of the histogram is significantly larger than the standard deviation 
$\sigma$ of the Gaussian fit (see ``sigma'' of the inset box), 
consistent with the long tail of non-Gaussian outliers which appear.
The upper right panel contains a histogram of and a fit
to only the cleaned 
data set, and now the histogram RMS and 
$\sigma$ agree within a factor of two. 
The lower panel shows the fit to the cleaned DR9 sample.
From these histograms we can estimate accurate 
statistical and systematic errors in the photometric parallax distance to M67.
}
\label{fig:m67gmigauss}
\end{center}
\end{figure}

The upper left panel of Fig.~\ref{fig:m67gmigauss} shows the DR7 data for the 
M67 cluster with only the proper motion cut applied.  The upper right
panel shows the same field with the unresolved binaries removed, as per
the ``cleaning'' procedure we have described, 
noting the
changed scale on the x-axis.  
The lower panel is the DR9 version of
the upper right panel, i.e., the binaries have been removed via a similar
procedure and the same
proper motion cut has been applied.  Each panel shows both a histogram of
the distances to the stars in the cluster as well as the results of a
Gaussian fit.
The fit parameters are listed in small
boxes at the upper right of each panel.  We list from the top down: the 
number of stars, the histogram mean and root-mean-square (RMS), 
as well as the information for the Gaussian fit, for which 
$f(x)= p_0 \exp(-(1/2)((x- \langle x \rangle )/\sigma)^2)$. That is, 
we show the $\chi^2$ per degree of freedom, 
as well as the fit parameters, namely, 
a constant ($p_0$), mean ($\langle x \rangle$), and 
sigma ($\sigma$). 
The two most relevant parameters to compare are 
the histogram RMS with the sigma from the Gaussian fit. 
These parameters should agree in the limit of a large number of points
if the data set is indeed normally distributed \footnote{http://root.cern.ch}. 
In particular, when a data set has many outliers, the histogram
RMS width may be very different from the Gaussian width.  For instance, in
the upper left panel, the histogram RMS is $0.81$, whereas $\sigma$ is only 
$0.11$.  After cleaning, the histogram RMS becomes $0.15$, whereas 
$\sigma$ becomes $0.08$ --- the difference has shrunk from a factor of 
eight to a
factor of two. 
 Looking at the DR9 sample, after a similar cleaning procedure has
been applied, the RMS is $0.2$, whereas 
 $\sigma$ is $0.05$, yielding a comparison 
somewhat worse than the fit to the DR7 data.
We are not concerned by this because the 
DR9 data set has only 65\% as many stars as the DR7 sample.  
We regard the 
difference between the histogram RMS and the Gaussian sigma as an estimate
of systematic error, reflective of the presence of non-cluster members, 
whereas sigma itself provides a statistical
error estimate. 
We combine these two types of error in quadrature,
using the results of the upper 
right panel of Fig.~\ref{fig:m67gmigauss}, to determine $\rm
d[{\rm M67}] = 0.89\pm 0.08(stat)\pm0.07(sys)$, reproducing the 
\citet{schlesinger12} distance by construction, to find 
a total cluster distance error of about 0.11 kpc. For the DR9 set, 
$\rm d[{\rm M67}] = 0.93\pm 0.05(stat)\pm 0.15(sys)$ kpc 
for a total error of 
0.16 kpc; 
the two data sets are clearly compatible. The latter result thus also compares 
favorably with the average literature value
we have determined from \citet{schlesinger12}, namely, 
$d_{\rm M67} = 0.89\pm0.06$ kpc. This gives us confidence
that we can employ our DR7-tuned  photometric parallax relations
on the DR9 data set. Note, too, we are able to assess the cluster distance
to within ${\cal O}(10\%)$. 
Noting once again the upper right panel 
of Fig.~\ref{fig:m67gmigauss}, we can translate the statistical 
error in the distance to the cluster to a typical error in the
distance to a star by multiplying 0.066 by the square root of the
number of cluster members, to yield an error of about 0.11 kpc. 
Consequently, we claim that we can assess the distance to a star
in M67 to ${\cal O}(10\%)$ as well.

\begin{figure}
\begin{center}
\includegraphics[scale=0.45]{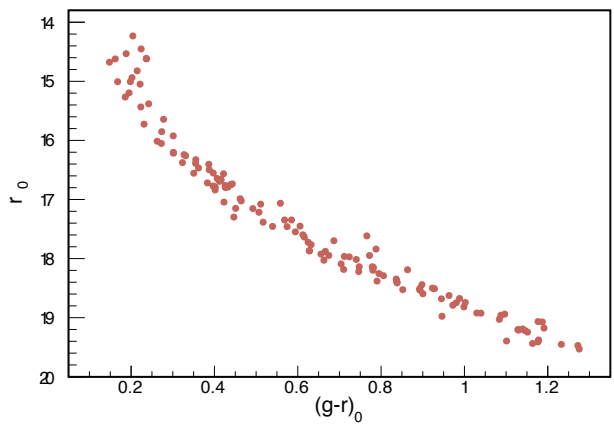}
\caption{
 Color magnitude diagram of stars in the NGC 2420 field with 
radial velocity close to that of the known cluster value.  
Nearly all outliers are removed by this procedure. 
}
\label{fig:ngc2420_hr}
\end{center}
\end{figure}
\begin{figure}
\begin{center}
\includegraphics[scale=0.55]{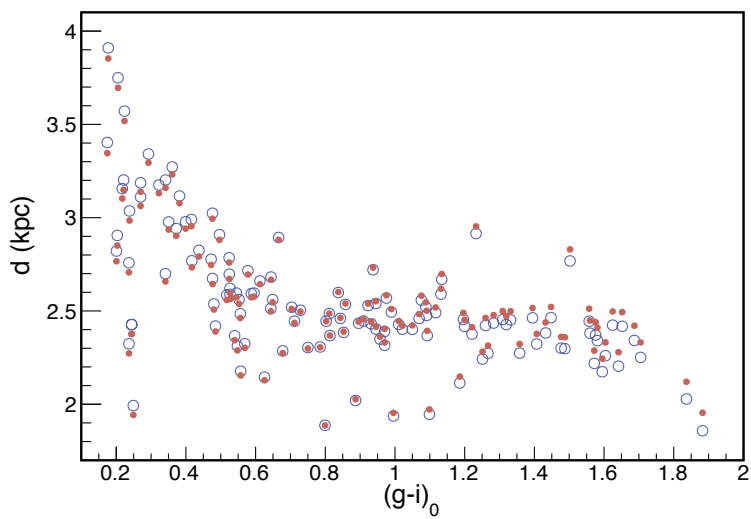}
\caption{
Photometric 
parallax relation of 
Fig.~\ref{fig:m67gmifix} applied to NGC 2420 
candidate members, after applying the radial velocity cut (see text). 
Open circles are the distances after applying Eq.~(\ref{ppgmi0}). 
The filled points represent the NGC 2420 stars after the same 
correction that was derived for M67, to yield 
the upper right panel of Fig.~\ref{fig:m67gmifix}, is applied here. 
Note that we use $\rm [Fe/H] = -0.37$ for the metallicity of the 
 NGC 2420 stars.}
\label{fig:ngc2420_dgmi}
\end{center}
\end{figure}
\begin{figure}
\begin{center}
\includegraphics[scale=0.55]{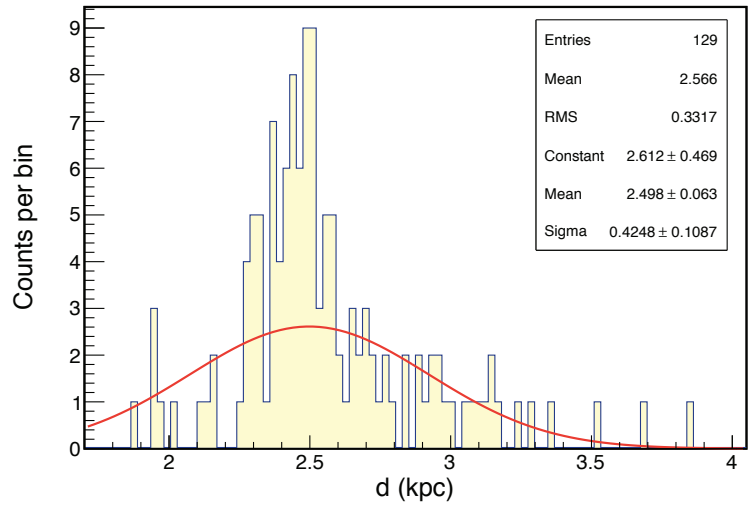}
\caption{
A photometric parallax histogram and 
fit to the DR9 NGC2420 candidate data set. 
The histogram RMS and Gaussian sigma are comparable, 
though both are broader than for the M67 data, 
suggesting 
the presence of non-cluster members.
}
\label{fig:ngc2420_dgmi_fit}
\end{center}
\end{figure}

We now turn to the analysis of 
the cluster NGC 2420, for which we employ a metallicity 
of $\rm [Fe/H] = -0.37$ \citep{antwarog06,lee08b}, though other
determinations have yielded a range of metallicities \citep{lee08b}. 
Figure \ref{fig:ngc2420_hr} shows a CMD of members of the 
cluster NGC2420 from DR9,  
which have further been culled by only keeping objects with 
SDSS heliocentric radial velocities compatible with 
the radial velocity of the cluster, 
$v_r = 75.1 \pm 5.9 \, \rm km\>s^{-1}$ \citep{lee08b}, though 
\citet{lee08b} note previous radial velocities measurements ranging from  
$67 - 84 \,\rm km\>s^{-1}$.  
We note that the systematic proper motion of NGC 2420 and that of
nearby Galactic field stars do not differ significantly, so 
that we cannot use
the method we employed for M67 to isolate cluster members here.
Velocities from 
SDSS stellar spectra typically have errors of $5 \, \rm km\>s^{-1}$, 
so that we do not restrict the cluster membership based on the 
error in the determined radial velocity and demand only that the
velocity itself fall within the interval 
$65 - 85 \, \rm km \>s^{-1}$. Since our ability to 
determine cluster membership is less crisp in this case, we do 
not try to refine our distance assessment further. Rather, we wish only 
to apply our present procedure to the stars of NGC2420, as an additional test. 
We show in Fig.~\ref{fig:ngc2420_dgmi} 
the photometric parallax relation for NGC2420 stars 
with (small filled circles) and without (open circles) the
slope correction derived from M67. The slope correction lessens
the variations in the computed distance with color slightly. 
In Fig.~\ref{fig:ngc2420_dgmi_fit} we compare 
the Gaussian fit to the histogram of NGC2420 distances. 
We apply the same procedure we used previously for M67 
to determine the systematic 
and statistical distance errors, finding 
$\rm d(NGC2420)= 2.5\pm 0.4(stat)\pm 0.09(sys)$ kpc for a total distance 
error of 
$0.41$ kpc. 
An error-weighted 
average of the distances given in \citet{schlesinger12}, assuming
the errors are independent, 
yields $d_{\rm NGC2420}=2.57 \pm 0.13$ kpc. 
%
The determinations compare favorably, though the width of the
Gaussian fit in this case is much broader than it was for the
M67 cluster. This could be reflective of the presence of a
range of metallicities within the cluster, giving a basis for
the range of metallicities found in studies of selected stars \citep{lee08b}. 

We now repeat our $(g-i)_0$-based analyses 
with one based on $(r-i)_0$ color 
and its associated photometric 
parallax of Eq.~(\ref{pprmi0}), after \citet{juric08}.
By so doing we wish to show that our 
assessment of systematic errors is independent of the particular 
choice of 
photometric parallax relation. This is also crucial for our ultimate 
application of these refined relations to the stars of our 
asymmetry analysis, for we wish to demonstrate that the results
which emerge there are also independent of the photometric relation
chosen. For this to be a sensitive test, the two relations must be
similarly well calibrated. 
Thus we have repeated the analysis depicted in 
Fig.~\ref{fig:m67gmifix} using the photometric parallax relation
in $(r-i)_0$ color. 
The determined linear correction in color
turns out to be slightly larger in this
case, yielding a corrected distance $d$, namely, 
\begin{equation}
d
 = d[r_0, M_r( (r-i)_0,[{\rm Fe/H}])]
+ 0.1415 (r-i)_0 + 0.0436 \,.
\label{drmi0tune}
\end{equation}
We have also confirmed that the 
same correction applied to the distances to the stars in 
the DR9 data set also yields a reasonably flat distance relation with color,
rather comparable to that shown in the third panel of 
Fig.~\ref{fig:m67gmifix} for $(g-i)_0$ color. 
While there are star-by-star 
differences between DR7 and DR9, 
the corrected photometric parallax 
relation remains flat --- no additional adjustments are needed, and
we do not make any. 
We have made Gaussian fits of the distances to M67, mirroring 
Fig.~\ref{fig:m67gmigauss} 
for 
$(r-i)_0$ color.  
For DR7 we obtain 
$\rm d(M67)_{(r-i)_0} =0.88\pm 0.13(stat)\pm 0.04(sys)$, 
whereas for DR9 we find 
$\rm d(M67)_{(r-i)_0} = 0.89\pm0.10(stat)\pm 0.10(sys)$, 
very comparable to the $(g-i)_0$ results.  
We have also studied the 
NGC2420 $(r-i)_0$ 
distances before 
and after 
applying the M67-derived correction to the NGC2420 candidate
members after the radial velocity cut has been applied. 
In this case the distance refinement yields at best a modest 
improvement.  Finally, 
the DR9 NGC2420 sample, refined by the radial velocity cut we have described, 
is fit, 
which yields $\rm d(NGC2420)_{\rm (r-i)_0}=2.4\pm 0.5(stat)\pm 0.2(sys)$.
This distance is also compatible with the reference distance
to this cluster. 

We conclude that our distance estimates, as well as their precision, 
are not sensitive 
to the specific photometric parallax relation we choose. 
We demonstrate that we are able to recover ${\cal O}(10\%)$ statistical 
distance errors for the cluster for which we have the
best membership information.  Moreover, we are able to 
assess the systematic errors incurred quantitatively by our 
assumption of cluster membership and find that they are no 
larger than our assessed statistical error. 
We have employed both parallax relations in the analyses
which follow, and find no marked differences, 
though we 
focus on the results 
determined through use of the $(g-i)_0$ color relation in what follows. 
We used a $(r-i)_0$ relation in W12; a comparison of 
the detailed calibrations 
reveal, however, that the correction required
to yield a M67 cluster distance independent of color is smaller
in the case of the $(g-i)_0$ relation. Moreover, we find that 
we can study a somewhat larger interval in $z$ in the $(g-i)_0$ case. 
Thus we favor this choice in what follows. 

\subsection{Stellar Identification} 
Although the great preponderance 
of the stars in our sample are
K- and M-type main-sequence stars, it is possible 
that giants can appear in our sample as well. 
Moreover, although their distances are 
reasonably 
determined by treating them as single, resolved objects, 
near-equal-mass unresolved binaries can also be present --- the distances
to the latter, as we have noted, 
are not well-treated by this approximation.
We consider each of these possibilities in turn. 

\begin{figure}
\begin{center}
\includegraphics[scale=0.55]{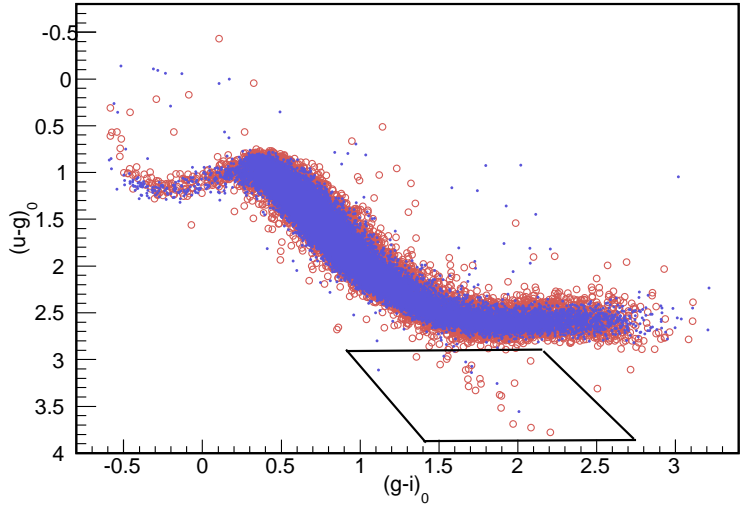}
\caption{Color-color diagram of $(u-g)_0$ versus $(g-i)_0$ to
identify non-main-sequence stars. We compare the population 
north and south: $b>0$ (northern) stars are small, filled circles, whereas 
$b<0$ (southern) stars are open circles. 
Here we only include objects with a
$u$-band error of less than $0.05$ 
and with $14.9 < r_0  < 15.4$.  
This selection favors brighter stars and thus 
giants over dwarfs.
A fainter selection in $r_0$ would reduce the fraction of giants even 
further. 
Nearly all the stars in this plot are dwarfs; however, giants do appear
and can be seen as a faint tail (mostly Sagittarius stream K/M giants)
extending 
out of the dense stellar locus with $(g-i)_0 > 1.4$ and $(u-g)_0 > 3$.
Giants are a completely negligible contaminant 
to our asymmetry analysis, for which $15 < r_0 < 21$ (see text).
}
\label{fig:giants}
\end{center}
\end{figure}

Our selection $13 < r_0 < 21$ is numerically dominated by 
faint-end-magnitude stars; this 
ensures that
giants can only play a small role in 
our sample. Exploiting the full multi-band nature of the SDSS data
also permits us to make an 
explicit numerical bound on their appearance. 
Figure \ref{fig:giants},
similar to Fig.~10 of \citet{yanny09b},
reveals the extent to which giants appear. 
Giants are visible in color-color space 
as a small diagonal tail running 
from $((u-g)_0,(g-i)_0) = (3,1.4)$ to $(4,2.2)$. 
Their numbers are completely swamped by dwarf stars; specifically
for the $r_0$ band selection $14.9 < r_0 < 15.4$, we
note that there are only some 30 giants in comparison to roughly
70,000 main-sequence stars, for a pollution of only some $0.04\%$. 
Only a fraction of the few 
giants which do appear in Fig.~\ref{fig:giants} can appear 
in the sample which we analyze in
this paper. 
Consequently we conclude that giants 
%
have no impact on either the number count asymmetries of W12 or
on the asymmetries we present here. 
Fig.~\ref{fig:giants} also splits the stars 
into solid circle (blue) ($b>0$)
and open circle (red) ($b<0$) subsamples. 
The distributions of these 
two samples in color-color space, while not identical, 
lie on top of each other, suggesting there 
is no gross stellar population difference, 
i.e., from either  metallicity or age, 
in 
these two sub-populations.  
We return to this point in the next section. 

We now turn to the possibility of unresolved, equal-mass binaries 
in our sample. 
Figure \ref{fig:hr_m67} shows a locus of stars shifted about 
0.7 mag brighter than the single star locus for the cluster M67; it
seems that some 10\% of the stars are near equal mass unresolved 
binaries \citep{fan96}. 
In the general field, we note the study of solar-like stars using
data from {\it Hipparcos} \citep{raghavan10}: 
Fig.~16 of that reference shows that 17
of 110 binaries have a mass ratio of 0.9 or greater, which is 
consistent within statistics with our study of M67. 
Even though we do not correct for this effect in the field,
there is nothing different about the way these objects affect apparent 
star magnitudes above or below the plane, and again, 
as with the contamination by giants, 
unresolved binaries cannot not alter the asymmetry results of 
W12 significantly.

\subsection{Role of Unmeasured Properties: Metallicity and Age Effects}
How do unmeasured properties such as age and metallicity impact 
the assessed distances to our stars?
Stellar age is a particularly straightforward issue for us, 
since we are dealing 
nearly exclusively with stars (K and M dwarfs) which have not yet begun 
to evolve off the main sequence. A spread in the ages of these stars will
not have a significant effect on their colors until they do so. 
Moreover, the impact of very young stars is also not an issue. 
The minimum height above the plane for our stars is no less than 300 pc, so that 
we are also confident that the vast majority
 of our stars ($>99\%$) are already on the main sequence and have
standard color-magnitude relations for main sequence K and M 
stars \citep{juric08}. 
T Tauri stars and other young stars which 
        may still be on their descending Hayashi track to the main sequence
        are restricted to distances much closer to the plane. 
In particular, the scale 
	height of molecular gas, where essentially all stars are born
	is only 74 pc \citep{dame87}.
%
The role of metallicity is a more difficult issue. 
Attempts to determine metallicity from 
photometry only \citep{ivezic08} almost invariably 
rely 
on the determination of 
$u$-band-connected colors because such colors 
are the most sensitive (for SDSS filters) to 
the presence or absence of the line-blanketing effects of UV atomic 
metal transitions. 
While we have access to the $u$-band magnitudes 
and $u-g$ colors for our stars, we 
do not have sufficient precision in the $u$ band to develop
a sufficiently accurate photometric proxy for the metallicity. 
Nevertheless, 
our Fig.~\ref{fig:giants}, with its closely overlapping North and South 
dwarf stellar loci, crudely demonstrates the absence of a 
strong shift in the 
average metallicity of the two samples. 
Although vertical and radial gradients in the metallicity are
known \citep{chen11}, the existence of 
possible north-south differences have not been documented. 
Certainly our Fig.~\ref{fig:giants} offers a simple probe of this issue
and suggests that it is unimportant, 
allowing us to set it aside. 


\subsection{Testing Dust Corrections} 


A remaining systematic effect concerns the possibility of 
inadequately estimated reddening differences between the 
North and South. 
While the \citet{schlegel98} dust maps have been shown, using photometry 
and CMD histograms, to be very accurate \citep{schlafly10}, 
especially for higher latitude ($|b| > 30^\circ$) 
sight-lines, there 
remains their finding, which is affirmed with a separate analysis using stars
with SEGUE spectra \citep{sf11}, 
that 
the stars in the South, working in $(g-r)_0$ color and using 
a very large scale average, 
appear to be about 0.02 mag redder 
than the stars in the North. 
This result is based on the average colors of blue turn-off 
stars. The difference in $(r-i)_0$ color 
is smaller by a factor of three.

There are three 
possible explanations for the observed color difference. 
First, there may be a global calibration error 
in the photometry within the SDSS DR7 (and DR8) releases, which 
the study of  \citet{schlafly10} supports, though 
this, noting Table 4 of \citet{schlafly10}, does not seem to 
capture the complete effect. However, the differences in area in the two studies
implies that the conclusions of \citet{sf11} need not hold for the photometric 
study \citep{schlafly10}. 
An offset in the $g_0$-band magnitude scale is the most likely 
because the blue-tip stars in $(r-i)_0$ color are  relatively 
unaffected. 
Maintaining absolute calibrations to better than 1\% 
across very large areas, especially those which are 
 only nominally contiguous, as it is here with different 
regions connected only by narrow strips running through 
the Galactic plane, is an extremely difficult task, 
subject to many subtle problems.  A recent analysis 
by \citet{betoule13} re-examines the large scale 
calibration of the SDSS with an effort toward 
improving Type I supernovae (standard candle) distances 
and finds that there are indeed systematic 
drifts in the SDSS calibration 
of some filters at the level of 1\% to 2\% 
over 100$^\circ$ scales. 
Thus we regard it as very possible that such 
a large-scale calibration drift exists in the 
$g_0$ band between the North and South.
A second way to explain a global color difference 
would be a reddening map anomaly.  
All our colors and magnitudes are corrected 
for reddening based on \citet{schlegel98}. 
We regard a problem with the reddening maps 
to be unlikely, as we discuss below, 
but we cannot completely rule out this
possibility. 
A third way to have a global color difference would be 
through 
an actual stellar population difference, namely, that 
the distribution of stellar
metallicity is different above and below the 
Galactic plane. This, too, we think is unlikely, and 
we discuss it further later.

\begin{figure}
\begin{center}
\includegraphics[scale=0.55,angle=-90]{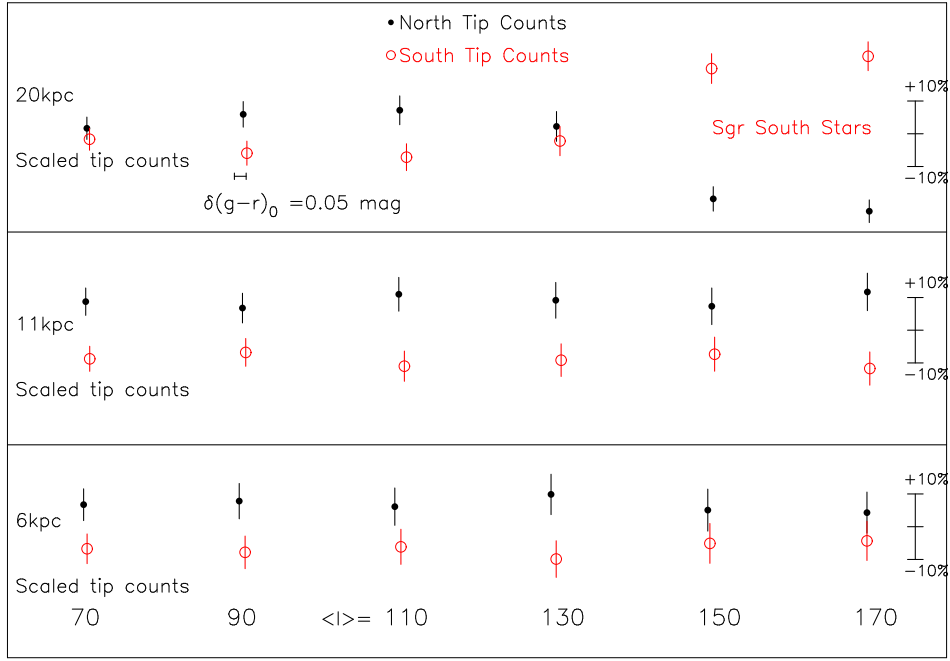}
\caption{A graphical representation of Table 1, the fraction of blue-tip 
stars in the North and South with $g_0$-band and $l$ selection.  The top row
represents the most distant stars, having $20 < g_0 < 21$ (d$\sim $ 20 kpc), 
and the
error bars on each point are sqrt($N$) number-count statistics, 
and the points themselves are normalized by the sum of counts, North
and South for that $g_0$- and $l$-band selection 
(see the scale bar to the right for the fractional differences in
counts).  The middle panel is for stars with $19 < g_0 < 20$, and the
lowest are for stars with $18 < g_0 < 19$, which are the nearest stars.  
Note in particular
the horizontal shift between North-South pairings as one moves from 
left to right.  This shift is the
average shift in the peak of the blue-tip in $(g-r)_0$ color, the
small horizontal scale bar in the top panel indicates  a 5\% shift.
}
\label{fig:bluetip}
\end{center}
\end{figure}

In order to select among the possibilities, 
we divide the 
``blue-tip'' stars (turn-off stars of spectral type F 
extending to the halo of the Milky Way), in our survey region into 
six 20$^\circ$ segments in Galactic $l$.  
We plot the $(g-r)_0$ color histograms of 
these blue-tip stars in three magnitude 
bins, namely $18 < g_0 < 19, \> 19 < g_0 < 20$, and $ 20 < g_0 < 21$, 
and measure the central color and peak heights of
each histogram in the north and south samples separately. 
Since \citet{schlafly10} uses $((g-r)_0, g_0)$ CMDs 
for their analysis of 
blue-tip stars, we adopt the same analysis for consistency. 
The results of this analysis are reported in Table 1, and the
number count studies and color offsets are also summarized 
in Fig.~\ref{fig:bluetip}. 
Table 1 presents peak $(g-r)_0$ turnoff color for star subsets, namely for
$50.3^\circ < |b| < 59^\circ$ and the indicated $l$ range
with $g_0$ restricted to the range given in brackets.
The numbers in parentheses indicate the total star counts
in the central bin. 
Note that the number counts of blue-tip stars in each bin should 
roughly match in the north and south (within Poisson errors), 
provided the halo is fairly smooth. 
Within Poisson statistics,
only two bins (in boldface) are markedly
discrepant --- and we will find that they correspond to the angular location
and distance of the turnoff stars in the Sagittarius stream.

\begin{figure}
\begin{center}
\includegraphics[scale=0.55]{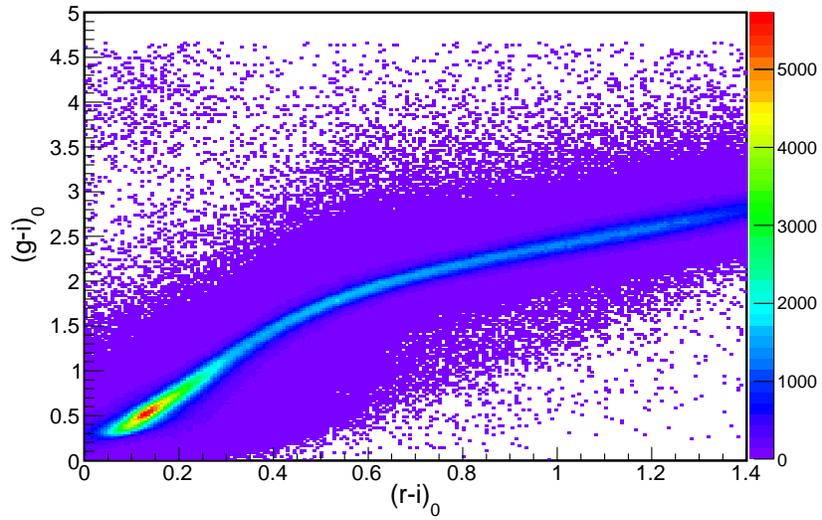}
\caption{
A $(r-i)_0$ versus $(g-i)_0$ color-color diagram for our stars ---
the smooth, tight correlation between the colors 
not only implies 
that the two different photometric parallax relations 
should be well-behaved but also suggests 
that they should give similar distance errors.
}
\label{fig:gmirmi}
\end{center}
\end{figure}

Generally Fig.~\ref{fig:bluetip} and Table 1 
show that the number of 
turn-off stars
tends to fall more or less smoothly for a fixed band of $g_0$ as 
$l$ increases, i.e., as one looks away from the Galactic center. 
There is one pair 
of bins in the South, however, 
with $20 < g_0 < 21$ and $140^\circ < l < 180^\circ$, 
which is strikingly inconsistent with that trend. In this case 
the difference in the number counts in matching bins North and South 
are several times the square root of the counts in the corresponding bins 
--- suggesting the presence of a structure.  In fact these 
bins in the South are contaminated by Sagittarius stream stars, 
as can be seen by an examination of Fig.~9 of \citet{yanny09a},
which clearly shows the Sagittarius stream passing through 
this $l$ range. The implied distances to 
the blue-tip stars, that is, $15 < d < 25$ kpc,  
closely 
matches that of the distance to the Sagittarius stream stars. 
Aside from this particular 
structure present in the South, 
the relative numbers of blue-tip stars in the North and 
South are fairly closely matched. 
We also note that the average colors of the 
South blue tip stars are systematically a couple of percent 
redder than the corresponding bins in the North, and these 
differences persist across four of the 20$^\circ$ longitude 
bins and three different magnitude bins, the latter corresponding
to three different distance bins ranging from some 5 to 25 kpc 
from the sun.  
This persistent color 
difference, which is just what \citet{schlafly10} found, argues 
strongly against the existence of a localized stellar population difference,  
or a localized ``rogue dust cloud,'' in either just the North or the South, 
but rather suggests 
a global calibration difference. 
In support of this 
we show in Fig.~\ref{fig:gmirmi} the comparison of $(g-i)_0$ and $(r-i)_0$ for 
stars in our sample. 
The tight central locus of Fig.~\ref{fig:gmirmi} which
appears without sharp bends, shows that one color may be substituted for the
other without large shifts of estimated (photometric parallax) distances. 
This in itself argues against the possibilities of rogue population or
reddening differences. 
Note that even with this global calibration difference, 
the star counts made in the North in comparison to those made in the 
South are not really affected because 
we see the same asymmetries 
whether we use the $(r-i)_0$ or the $(g-i)_0$ 
photometric parallax relation.
We have shown, too, 
by exploring photometric parallax 
relations in two correlated color ranges, 
$(g-i)_0$ and $(r-i)_0$, that our distance estimates are independent of the 
filters used.  


\section{The North-South Asymmetry}
\label{asym}

We now proceed to analyze our larger sample in the spirit of W12, 
to determine the vertical distribution of the stars and finally
the  North-South asymmetry they possess with respect to the center of the
Galactic plane. 
To do this, we must 
first determine the appropriate saturation and color cuts. 
We then construct a selection function in order to estimate the 
stellar number density as a function of 
vertical displacement from the sun, 
given that our stellar sample is restricted to a limited 
window in $b$ and $l$.
We note that with suitable 
color cuts the selection function is determined 
by purely geometric factors. Finally, we fit our inferred ``true''
vertical distribution of stars to a theoretical form in order to 
determine the location of the sun with respect to the Galactic plane. 
With this in hand we can finally determine the North-South
asymmetry about the Galactic plane 
--- and we have sufficient statistics to analyze its variation
with Galactic longitude as well. 

\begin{figure}
\begin{center}
\includegraphics[scale=0.45]{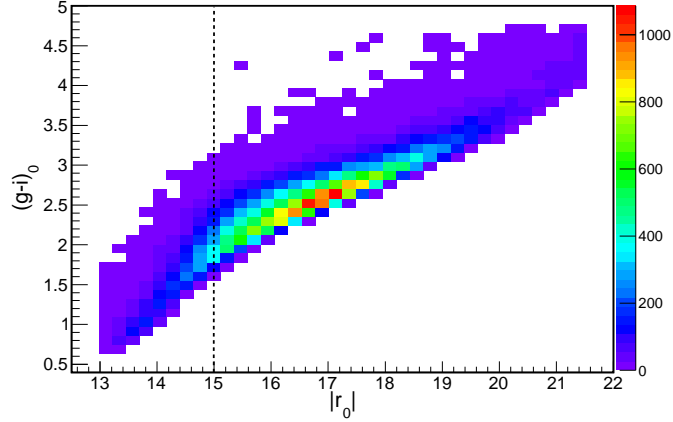}
\caption{
A $(g-i)_0$ versus 
$r_0$ CMD 
for stars with $|z| < 0.3$ kpc, computed as per
Eq.~(\ref{dgmi0tune}). 
Here we explore the appropriate bright-end limit
 of our data set, noting 
that saturation begins to impact some colors strongly for 
stars with $r_0 \sim 14.5$ and brighter.  
We set the saturation cut at $r_0 = 15$ (vertical line) 
and exclude all brighter stars, in order to make the 
bright-end limit more uniform in color. 
}
\label{fig:idgiants}
\end{center}
\end{figure}
\begin{figure}
\begin{center}
\includegraphics[scale=0.55]{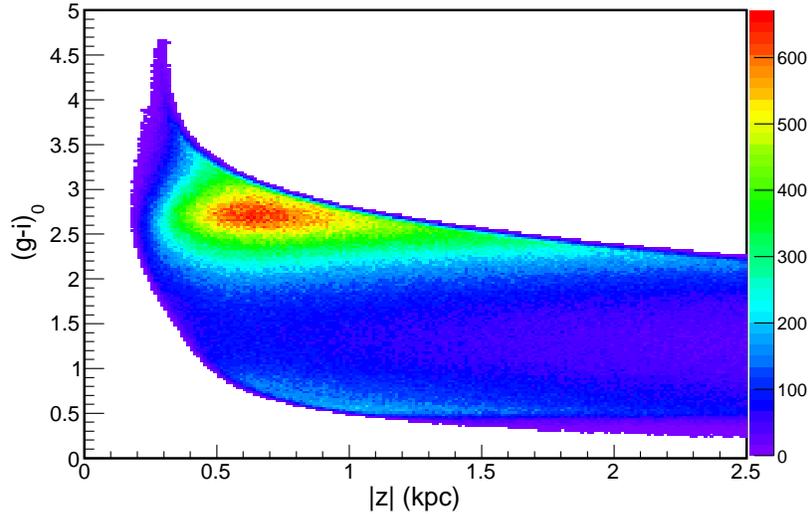}
\caption{
A color-distance diagram for our K/M dwarf sample. 
This figure of $(g-i)_0$ versus the 
absolute distance above the plane $|z|$ allows us 
to determine the range in $|z|$ over which a given 
band of $(g-i)_0$ color is complete 
(see text for the precise cuts used).
}
\label{fig:colorvsmetals}
\end{center}
\end{figure}
\begin{figure}
\begin{center}
\includegraphics[scale=0.60]{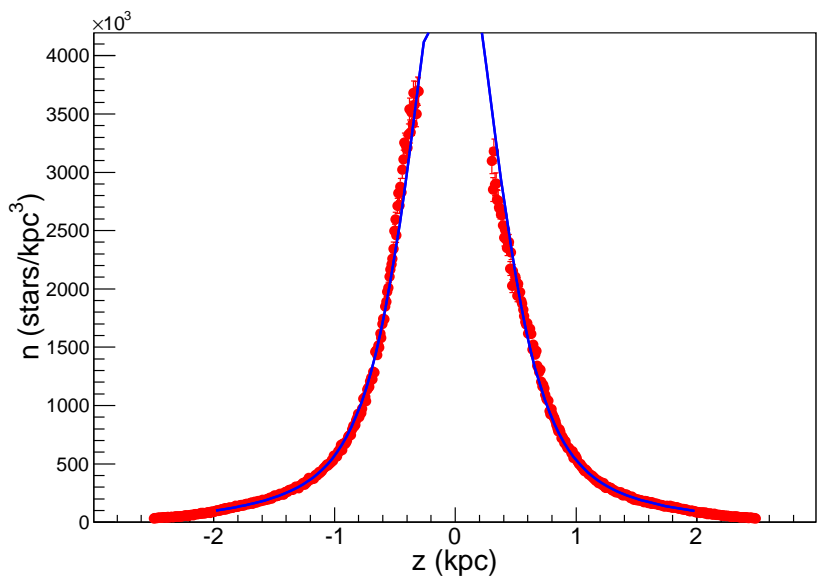}
\caption{
Histogram of the stellar density  as a function of the vertical displacement
$z$ from the sun. 
We employ the selection function and color and saturation cuts
described in the text and compute 
distances using Eq.~(\ref{dgmi0tune}). 
Even at this coarse scale, 
the wave-like North-South asymmetry can be seen on top of the symmetrical model.
}
\label{fig:fig1recap}
\end{center}
\end{figure}
\begin{figure}
\begin{center}
\includegraphics[scale=0.38]{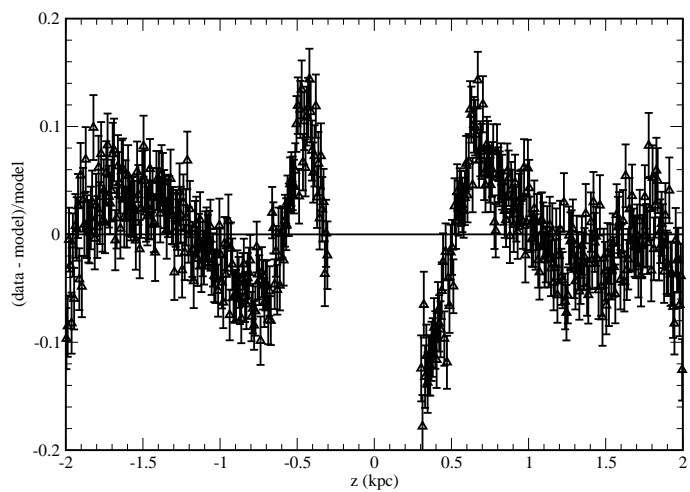}
\caption{
Difference between the estimated true stellar 
density and the best-fit model with vertical displacement from 
the sun, plotted
as $({\rm data}- {\rm model})/{\rm model}$, determined
from the inputs to Fig.~\ref{fig:fig1recap}. 
The residuals are dominantly of odd parity under $z \to -z$. 
}
\label{fig:oddparity}
\end{center}
\end{figure}

The SDSS data are affected by a selection on the bright end --- 
most objects are saturated if $r_0 < 14.5$.  In order to make
our saturation limit more uniform in color, we apply a 
slightly fainter saturation cut (though we used $r_0=14$ in W12). 
Its intended effect is to remove any 
systematics due to saturation in our asymmetry analysis.  
Fig.~\ref{fig:idgiants} shows that we may remove stars with 
$r_0 < 15$ --- this improves the completeness of the 
sample on the near end, without significantly 
losing statistical significance. 
Fig.~\ref{fig:colorvsmetals} shows how the 
particular $(g-i)_0$ colors of the stars 
relate to their vertical distances, where
we impose $r_0 > 15$ and compute distances as per Eq.~(\ref{dgmi0tune}). 
The shape of the region is determined by our window on $r_0$; a 
judicious choice of color interval opens a vertical window on $|z|$
which is not limited by our finite range of $r_0$. 
We note, e.g., that if we choose $1.8 < (g-i)_0 < 2.4$, our sample
is complete on the interval $0.3 \,{\rm kpc} < |z| < 2.0\, {\rm kpc}$.
In DR9, there are some 794,500 stars which satisfy all our cuts. 
The selection function is determined by geometry. Since we bin the
photometric data with vertical height, the selection function is the
effective volume, bin by bin, associated with the observed stars. 
That is, if the center of a vertical bin is at $z$, then 
the selection function ${\cal V}(z)$ for our North-South symmetric 
data sample is 
\begin{equation}
{\cal V}(z) =
\frac{1}{2} 
\delta (l_2 - l_1) z^2 \left(\frac{1}{\sin^2 b_1} - 
\frac{1}{\sin^2 b_2} \right) \,,
\end{equation} 
where $\delta$ is the bin width in kpc, noting  $l\in [l_1,l_2]$ 
and $|b| \in [b_1, b_2]$ are in radians. 
Finally we fit the function determined by 
$n(z)\equiv n_{\rm raw}(z)/{\cal V}(z)$, where $n_{\rm raw}(z)$ is
the number of stars we observe in a vertical bin centered on $z$. 
We adopt the same standard Galaxy thin and thick disk model form used in W12, 
\begin{equation}
n(z)=n_0 \left({\rm sech}^2\left(\frac{(z+z_{\odot})}{2H_1}\right) 
+ f {\rm sech}^2\left(\frac{(z+z_{\odot})}{2H_2}\right) 
\right) \,,
\end{equation} 
and show the result in Fig.~\ref{fig:fig1recap}, which displays 
the stellar number density as a function of distance. 
Already here we 
see that 
the deviations of the counts from the best-fit model 
change sign under $z \leftrightarrow -z$; i.e., they are odd under parity. 
This is further highlighted in Fig.~\ref{fig:oddparity}.  
We have 600 bins 
over the region $|z| < 2.5\,{\rm kpc}$, 
so that $\delta \approx 0.0083\,{\rm kpc}$. The fit is over the
region with $|z| < 2\,{\rm kpc}$, yielding 
$\chi^2/{\rm ndf} \approx 3.8$, $n_0 = (4.74\pm 0.04)\times 10^6\,{\rm kpc}^{-3}$, 
$H_1 = 0.232 \pm 0.001 \, {\rm kpc}$, $H_2 = 0.666 \pm 0.006\, {\rm kpc}$, 
$z_{\odot} = 14.3 \pm 0.6 \,{\rm pc}$, and $f=0.109\pm 0.002$. 
Figures \ref{fig:fig1recap} and \ref{fig:oddparity} 
update Fig.~1 in W12. The quality of our fit is markedly better than in W12, 
due to an improved calibration and saturation-cut choices, 
though we still reject a North-South symmetric model for $n(z)$. 
The results of Figs. \ref{fig:fig1recap} and \ref{fig:oddparity} 
evince a North-South asymmetry, strongly confirming 
the results of  W12.

\begin{figure}
\begin{center}
\includegraphics[scale=0.60]{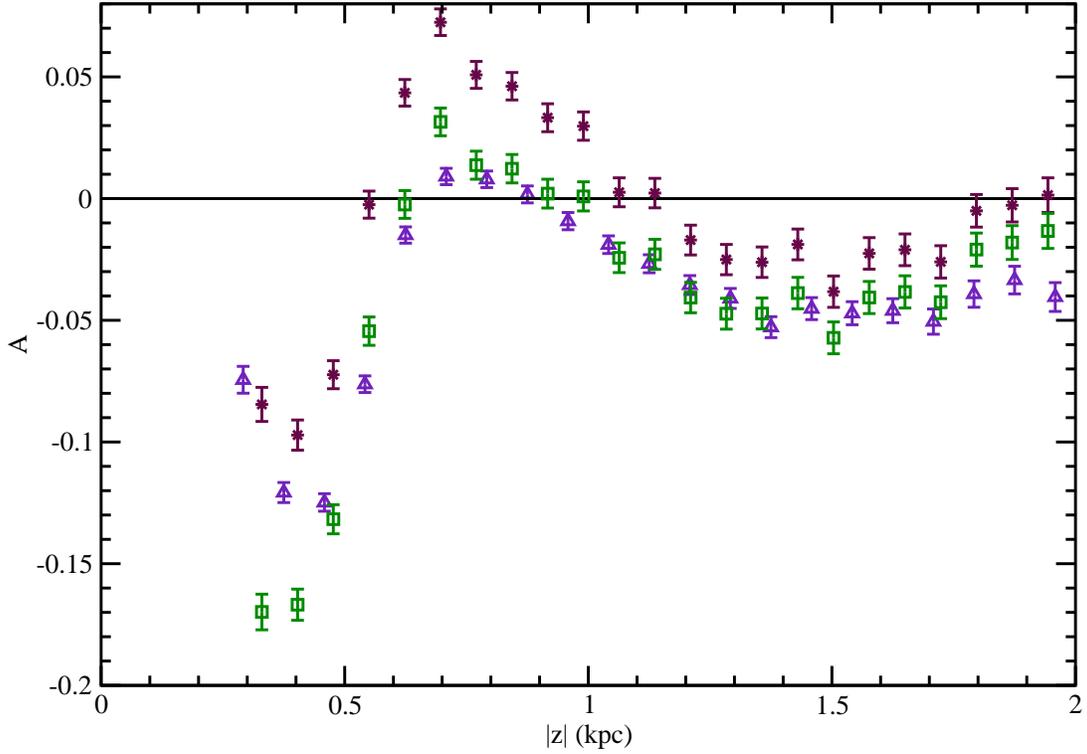}
\caption{
The asymmetry in star counts, North and South, with height $|z|$,
using $(g-i)_0$ parallax with $1.8 < (g-i)_0 <2.4$, through
stages of the analysis. 
The triangular (purple) points 
show the asymmetry in observed star counts,
${\cal A}_{\rm raw}(z)$ (see text), where $z=0$ in this case is
referenced to the sun's location. 
The square (green) 
points show the asymmetry in observed star counts
with $z$ replaced by $z + z_{\odot}$ so that 
the origin is now the center of the Galactic plane, though the points
are plotted on the same scale.
We have used $z_\odot = 14.3\,{\rm pc}$. 
The asterisk (maroon) 
points show our final North-South asymmetry ${\cal A}(z + z_\odot)$ (see text); 
here, too, the origin is the center of the Galactic plane. 
In this case the selection function has been applied as well. 
}
\label{fig:fig1recapx}
\end{center}
\end{figure}
\begin{figure}
\begin{center}
\includegraphics[scale=0.60]{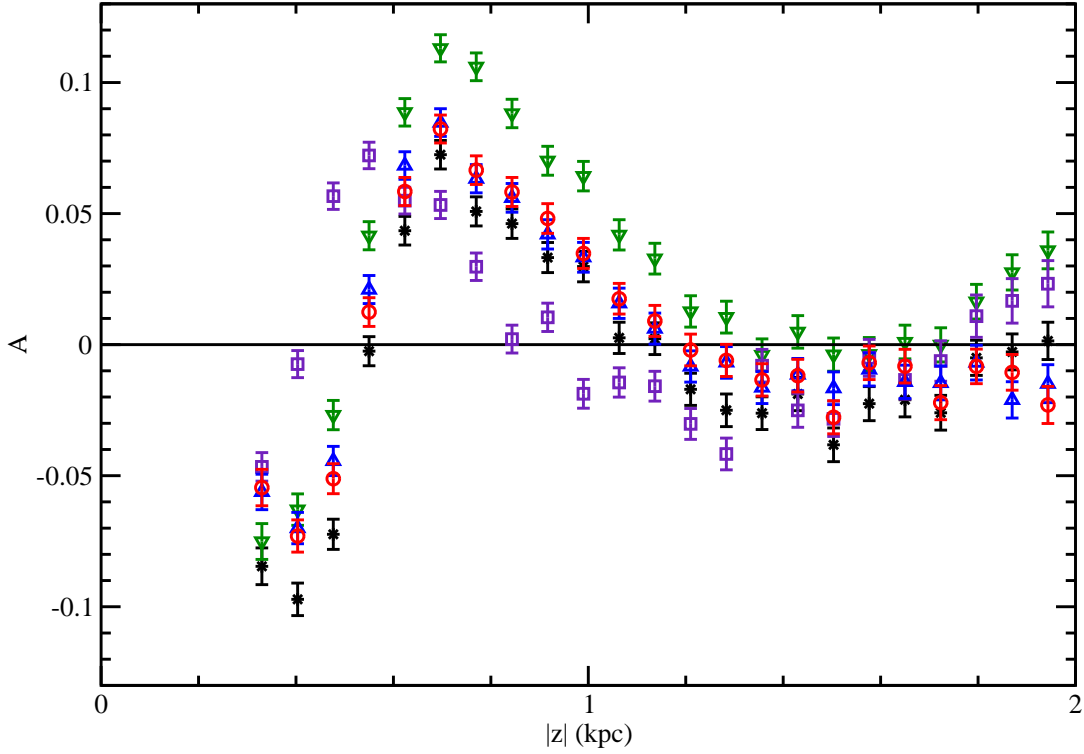}
\caption{
The asymmetry in star counts, North and South, with height $|z|$ from the
center of the Galactic plane, 
using $(g-i)_0$ parallax with $1.8 < (g-i)_0 <2.4$, upon the
variation of different systematic effects. 
The asterisk (black) 
points show our final North-South asymmetry 
of Fig.~\ref{fig:fig1recapx}, for which we use 
$z_\odot = 14.3\,{\rm pc}$, determined from the fit of
our star counts.  The down-pointing triangular
(green) points show the asymmetry which emerges if 
$z_\odot$ is changed to $30\,{\rm pc}$. The square (purple) 
points shows the asymmetry if the adjustment of
the photometric parallax relation from the fit of the 
M67 cluster, Eq.~(\ref{dgmi0tune}), is not employed. 
The last two curves illustrate the consequence of
a possible 2\% $g$-band calibration error on the
asymmetry. The up-pointing triangular (blue) points 
shows the asymmetry which emerges if $g \to g +0.02$ 
in the North only, and the circular (red) points illustrate
what happens if $g\to g -0.02$ in the South only. 
}
\label{fig:asym_fixes}
\end{center}
\end{figure}

We construct a North-South asymmetry by comparing star counts North and
South directly. If we compare, e.g., the raw counts as a function of
the vertical displacement 
from the sun's location, 
we define 
\begin{equation}
{\cal A}_{\rm raw} (|z|) \equiv 
\frac{n_{\rm raw}(|z|) - n_{\rm raw}(-|z|)}{n_{\rm raw}(|z|) 
+ n_{\rm raw}(-|z|)} \,.
\label{asymraw}
\end{equation}
This quantity is shown in Fig.~\ref{fig:fig1recapx}. 
Ultimately, however, we would like to determine the North-South
asymmetry in terms of the effective stellar number density 
as a function of the vertical displacement from the Galactic plane. 
We realize  
this in two steps. First, we plot 
the raw asymmetry in terms of $|z + z_{\odot}|$ 
as shown in Fig.~\ref{fig:fig1recapx}. We then 
include the selection function as well, in order to compute the asymmetry
in terms of the effective number density, namely, 
\begin{equation}
{\cal A} (|z+z_{\odot}|) \equiv 
\frac{n(|z+z_{\odot}|) - n(-|z+z_{\odot}|)}{n(|z+z_{\odot}|) 
+ n(-|z+z_{\odot}|)} \,.
\label{asymphys}
\end{equation}
We use $z_\odot = 14.3\rm \> kpc$ as per our earlier fit. 
The North-South asymmetry is visible in the original star
number counts, and the refinements we have included in 
Eq.~(\ref{asymphys}) impact its precise shape but not its
significance. At this point it is appropriate to revisit 
the various systematic effects we have studied earlier
in the paper and explore their impact on the asymmetry. 
Specifically, we consider the role of $z_\odot$, the
impact of our tune of the photometric parallax relation
from the fit to M67 data, Eq.~\ref{dgmi0tune}, and finally
the impact of a possible 2\% calibration error in the $g_0$
band, either North or South. The asymmetries which
result from these changes are 
illustrated in Fig.~\ref{fig:asym_fixes}. In all cases
the resulting asymmetry is still significantly non-zero. 
However, the impact of the change in $z_\odot$ and in the
tilt and offset of the photometric parallax relation away
from the optimal values determined in our fits yield significant
changes in the shape of our asymmetry. The impact of the 
possible 2\% calibration errors is, in contrast, much smaller. 

%

\begin{figure}
\centering
\includegraphics[scale=0.60]{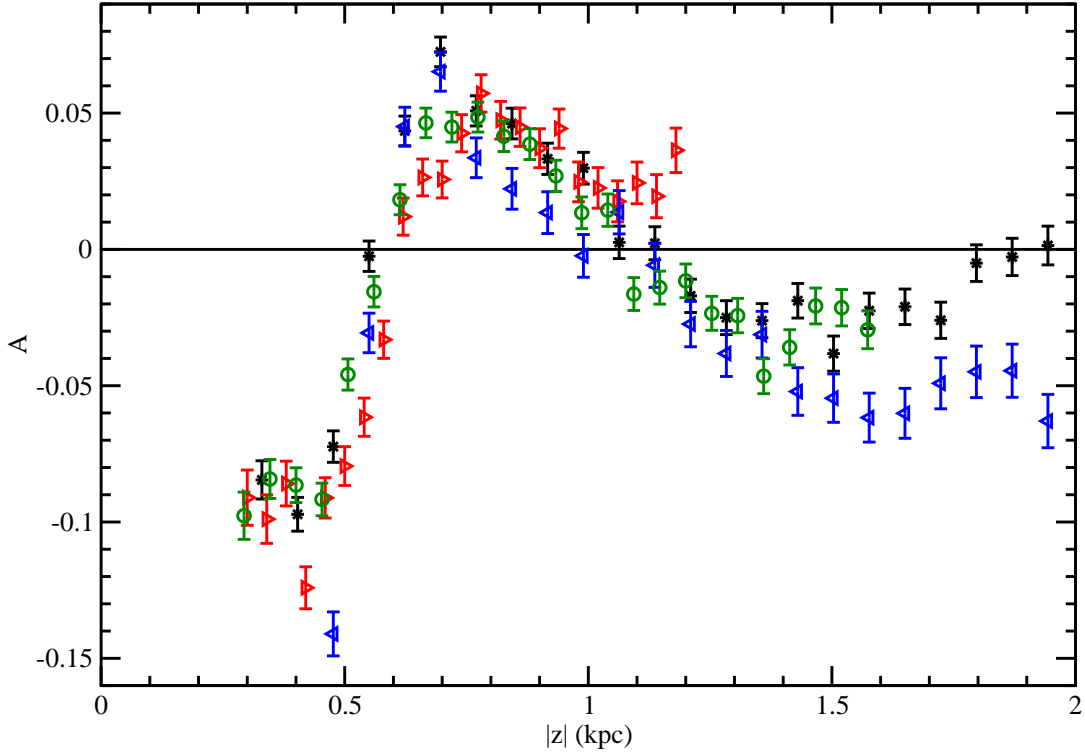}
\caption{
The asymmetry in star counts, North and South, with height $|z|$ from
the Galactic plane, 
for different bands in $(g-i)_0$ color. The asterisk (black) 
points
correspond to $1.8 < (g-i)_0 <2.4$; these points are the same
as the asterisk points depicted in Fig.~\ref{fig:fig1recapx}. 
The left-pointing-triangle (blue) 
correspond 
to $0.95 < (g-i)_0 <1.8$, whereas the right-pointing-triangle 
(red) points correspond
to $2.4 < (g-i)_0 < 2.7$. 
We
also include an asymmetry analyzed using our distance relationship
based on $(r-i)_0$ color, Eq.~(\ref{drmi0tune}); 
the circle (green) points correspond to 
$0.6 < (r-i)_0 < 1.1$. 
In each case the window in $|z|$ shown 
is determined by the $r_0$ saturation limits for the given 
color interval, as determined, e.g., for $(g-i)_0$ color from 
Fig.~\ref{fig:colorvsmetals}. 
We employ $z_\odot=14.3\,{\rm pc}$ 
throughout, and emphasize that $|z|=0$ here is referenced to the center
of the Galactic plane. 
}
\label{fig:fig2recap}
\end{figure}

In Fig.~\ref{fig:fig2recap} we compare the asymmetry 
${\cal A}(|z+z_{\odot}|)$ computed using different color bins; 
the window on the vertical coordinate 
for each color band is limited by the saturation
and faintness limits on $r_0$ in each case, 
noting, e.g., Fig.~\ref{fig:colorvsmetals} for $(g-i)_0$ color. 
We show the asymmetries for three different bins on 
$(g-i)_0$ color, namely, $1.40 < (g-i)_0 < 1.8$, 
$1.8 < (g-i)_0 < 2.4$, and $2.4 < (g-i)_0 < 2.7$. 
The different asymmetries are reasonably similar where
they all coexist.
We also compare these results with an asymmetry determined
from distances computed using $(r-i)_0$ color, namely 
Eq.~(\ref{drmi0tune}). Choosing a sample with 
$0.6 < (r-i)_0 < 1.1$, we note as per Fig.~\ref{fig:gmirmi} 
that this crudely corresponds to $1.8 < (g-i)_0 < 2.4$. Perhaps
not unexpectedly, the asymmetry results in these two 
particular bands compare favorably, 
though the stars selected by the $(r-i)_0$ cut are a distinct sample --- 
some 852,000 stars satisfy the cuts for this analysis. 
This particular result confirms, moreover, 
the findings of W12 with a sample size 
roughly three times bigger. Interestingly, the close comparison
of the results based on the different photometric parallax relations
suggests that any possible problem with the large-scale calibration
of the $g_0$ band does not impact the asymmetry significantly; 
this is supported by our study of Fig.~\ref{fig:asym_fixes} as well. 



\begin{figure}
\centering
\includegraphics[scale=0.60]{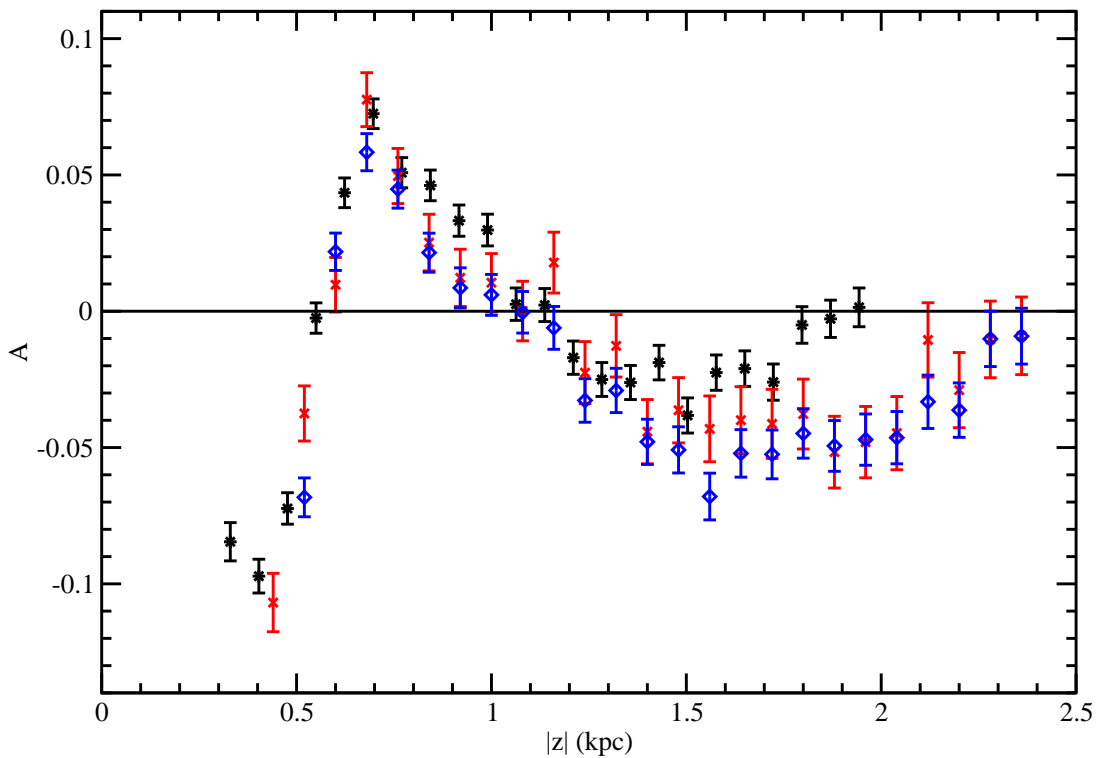}
\caption{
The asymmetry in star counts, North and South, with height $|z|$ from
the Galactic plane, 
for different bands in $(g-i)_0$ color. The asterisk (black) 
points
correspond to $1.8 < (g-i)_0 <2.4$; these points are the same
as the asterisk points depicted in Figs.~\ref{fig:fig1recapx}
and \ref{fig:fig2recap}. 
The diamonds (blue) 
correspond 
to $0.95 < (g-i)_0 <1.8$, whereas the 
crosses (red) correspond
to $1.4 < (g-i)_0 < 1.8$ --- some of the ``contamination'' in 
the $|z| \sim 1.5\,{\rm kpc}$ region is removed through this
more restricted color cut, but the data seem to suggest the
existence of a nonzero asymmetry at these heights as well. 
}
\label{fig:asymcutout}
\end{figure}

We consider now the persistently large asymmetry 
seen
in Fig.~\ref{fig:fig2recap} at larger distances from the plane, 
$1.4 < |z| < 2.0\, {\rm kpc}$ for the 
bluest color bin,
$0.95 < (g-i)_0 < 1.8$. The blue end of this bin corresponds to K-type stars
with $(g-r)_0 \sim 0.72$.  Could this excess be due to 
a contamination in this field? We note that 
giants in the Sagittarius stream's southern tail are located in our observing
window some 15-30 kpc below the plane. 
Figure 11 from \citet{m03} shows that 
only approximately 100 such M giants exist below the Galactic 
plane. Still, perhaps stars
associated with the more populous red giant branch and even sub-giant branch
stars associated with Sagittarius could add to the effect. 
By inspection of our sample, we note that the magnitude and color of the
largest excess of stars with $-2 < z < -1.5$ kpc have an average
$\langle r_0,(g-r)_0 \rangle 
 = 17.5, 0.63$.  These values correspond to a spectral class 
G/K sub-giant with absolute $M_r \sim 1.6$ (see Fig.3 of \citet{yanny09a}
for a view of the turnoff of the Sagittarius stream in the North 
at 45 kpc from the sun,
somewhat fainter than the magnitudes explored here), at a distance 
of about 15 kpc, close to the 15-30 kpc distance range for 
known Sagittarius stream south stars.
Are there enough of them?  For every M-giant (noting the $\sim1 00$ 
seen in \citet{m03}),
there can be 50 sub-giants in an evolving cluster luminosity function of
this age and metallicity, 
though not all of these will have colors
as red as $(g-i)_0 > 0.95$. Indeed 
most subgiant stars are bluer than our bluest
cut in Fig.~\ref{fig:fig2recap}. Still a significant portion of these stars 
are likely to remain. To explore the issue further, 
we study the asymmetry as a function of the bluest $(g-i)_0$ colors
we choose to include 
in Fig.~\ref{fig:asymcutout}. We see explicitly that restricting
the color from $0.95 < (g-i)_0 < 1.8$ to $1.4 < (g-i)_0 < 1.8$ 
does reduce the number of stars in the South at
$|z| \sim 1.5\,{\rm kpc}$. Nevertheless, this change has 
no impact and significance of the asymmetry we claim at smaller $|z|$, 
and some asymmetry in the region of $|z| \sim 1.5\,{\rm kpc}$ would
seem to persist. 
We have studied the 
asymmetry split into latitude bins as well, and it 
further supports this interpretation, 
with the finding of a similar asymmetry only in the bin containing
Sagittarius stream stars (see Table 1) at $140 < l < 180$ compared with the
lower latitude bin ($l < 140$), which shows a smaller asymmetry.
While the apparent third peak of the asymmetry remains 
visible in Fig.~\ref{fig:fig2recap} at the 3-4\% level when we 
exclude sub-giant contaminants by restricting our choice to 
redder stars $(g-i)_0 > 1.8$, (corresponding to a late K, early M
spectral type) we could have some contamination from Sagittarius giants
affecting the amplitude of the third peak at the 1\% level.
Conversely, there is no evidence that the two larger peaks at 
$|z| \sim 0.4,\, 0.8\, {\rm kpc}$ 
could be due to sub-giant contamination from a more distant stream.
The asymmetries are color-independent and significantly larger in
amplitude than any known stream could accommodate --- Sagittarius is the 
largest known Milky Way halo stream. 

\begin{figure}
\includegraphics[scale=0.6]{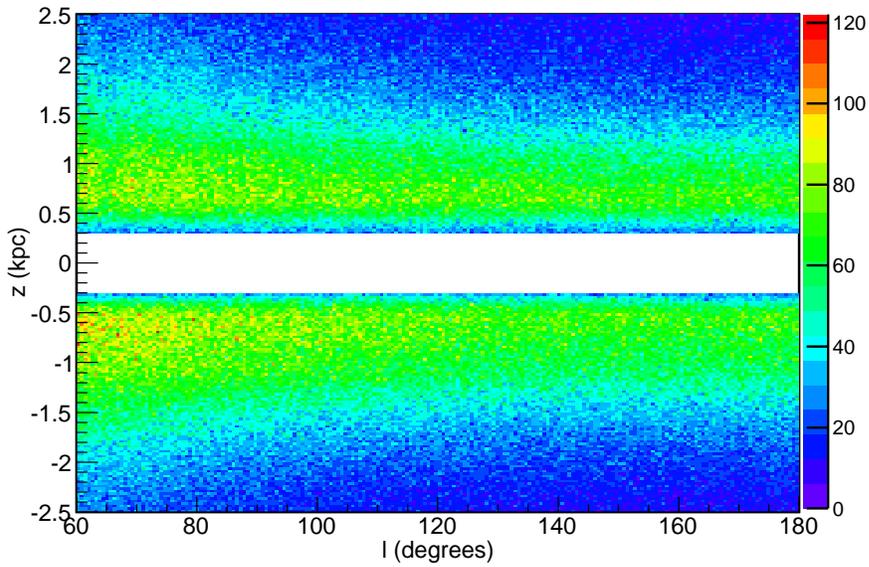}
\caption{
The stellar number counts with height $|z|$ from
the Galactic plane, using $(g-i)_0$ parallax with $0.95 < (g-i)_0 < 2.7$, 
and with $l$. We have imposed that $|r_0|>15$ and that 
$0.3 < |z| < 2.5$ kpc. This is the sample employed in our
analysis of the asymmetry in North-South number counts with $l$. 
}
\label{fig:zlsweep}
\end{figure}

\begin{figure}
\includegraphics[scale=0.6]{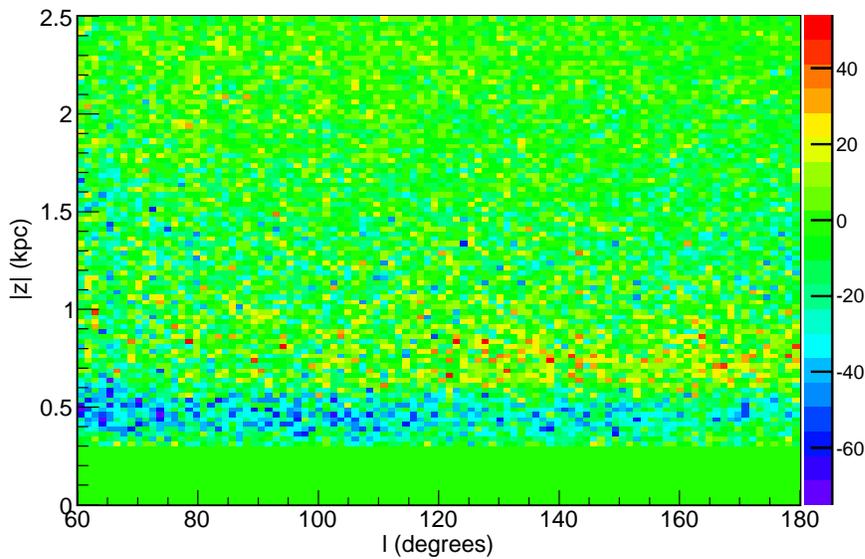}
\caption{
The difference in the stellar number counts, North minus South, 
 with height $|z|$ from
the Galactic plane, using $(g-i)_0$ parallax with $0.95 < (g-i)_0 < 2.7$, 
and with $l$. This is realized from the data displayed in 
Fig.~\ref{fig:zlsweep} and thus contains the same cuts. 
The features of the asymmetry presented in 
Fig.~\ref{fig:ldep} are already apparent in the raw data, as we
show here. 
}
\label{fig:zlsweep_diff}
\end{figure}

\begin{figure}
\centering
\includegraphics[scale=0.60]{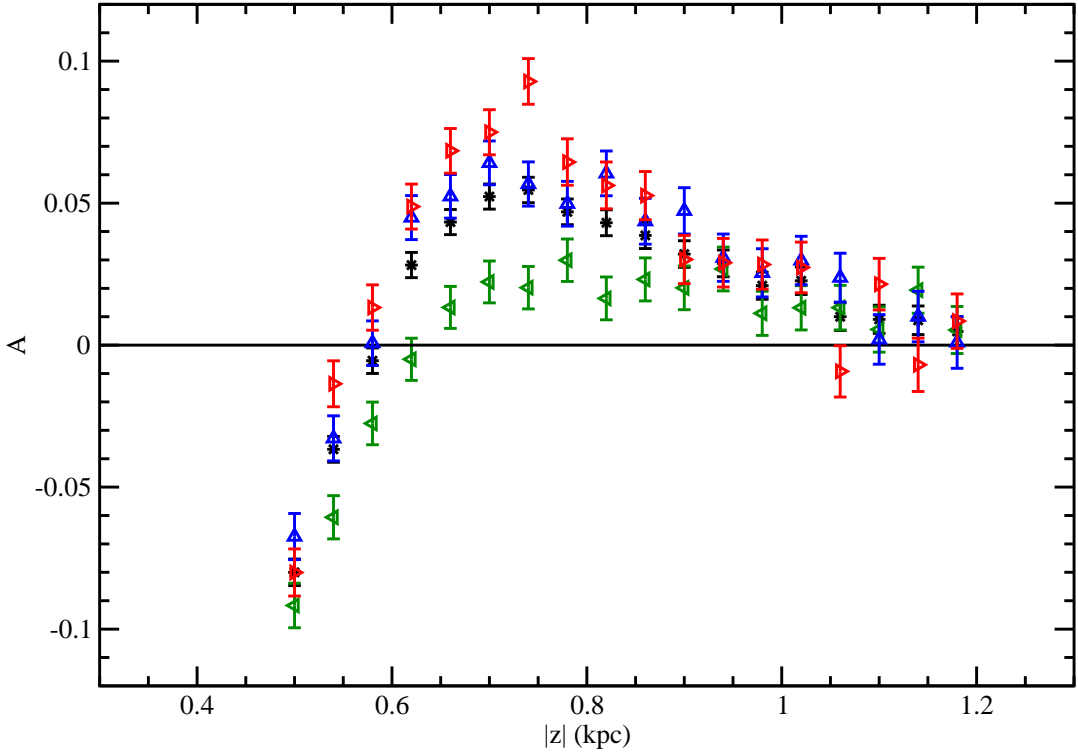}
\caption{
The asymmetry in star counts, North and South, with height $|z|$ from
the Galactic plane, using $(g-i)_0$ parallax with $0.95 < (g-i)_0 < 2.7$, 
for different ranges in $l$. 
The asterisk (black) points are the results for our complete range in 
$l$, $60^\circ \le l \le 180^\circ$. 
The left-pointing triangles (green) 
correspond to  $60^\circ \le l \le 100^\circ$.  
The up-pointing triangles (blue) correspond to  $100^\circ \le l \le 140^\circ$,
and the right-pointing triangles (red) 
have $140^\circ \le l \le 180^\circ$. 
As in Fig.~\ref{fig:fig2recap} we employ $z_\odot=14.3\,{\rm pc}$ 
throughout, while $|z|=0$ is referenced to the center
of the Galactic plane. 
}
\label{fig:ldep}
\end{figure}

Finally we consider the Galactic longitude dependence
of the asymmetry. Here we work in $(g-i)_0$ color, 
choosing the band $0.95 < (g-i)_0 < 2.7$. 
The analysis sample is shown in Fig.~\ref{fig:zlsweep};
it appears rather uniform, North and South. The number
counts differences, North and South, are also shown 
in Fig.~\ref{fig:zlsweep_diff}; here we see in the raw
data a persistent excess of counts in the South, and then
in the North, as we move out of the Galactic plane, for a
sweep of $l$. Finally, we report the determined North-South
asymmetry in Fig.~\ref{fig:ldep}. 
In the latter 
we restrict our consideration 
to the vertical window $[0.5 \,{\rm kpc}, 1.1 \,{\rm kpc}]$, the range
for which the sample is complete. For this larger color swatch
there are some 1.98 million stars in our sample. To study
the $l$ dependence, we split this sample 
into three forty-degree-wide $l$ bins. For 
$60^\circ < l < 100^\circ$, there are some 783,000 stars with a
mean $(g-i)_0$ color of $\langle (g-i)_0 \rangle =1.95$. 
For $100^\circ < l < 140^\circ$, there are some 635,000 stars with 
 $\langle (g-i)_0 \rangle =1.97$. 
For $140^\circ < l < 180^\circ$, there are some 566,000 stars with 
 $\langle (g-i)_0 \rangle =1.99$. 
The asymmetry for all $l$ 
in this color band is shown by the asterisk points in Fig.~\ref{fig:ldep}. 
We note that indeed there appears to be a significant difference 
in 
the asymmetry with $l$, with a smaller amplitude 
towards the Galactic center.
This is naively what one would expect for a  disk which gets 
thicker the closer one gets to the Galactic center; it would be 
less responsive to a disturbing force. Our forward- and
backward-looking samples, noting Fig.~\ref{fig:zvsR}, probe regions of the disk
separated by roughly 1 kpc. 
The longitudinal angle dependence
could also be suggestive of 
the location of a perturbing influence. 
In a followup 
paper (Gardner et al., in preparation), we plan to examine 
how well-localized the asymmetry may be.
We know that there are also effects due to a bar toward the Galactic
center which are not completely symmetric north and south, left and
right \citep{benetal05}, though such effects do not figure in our
present study. 

\section{Summary and Outlook}

In the context of a larger data set we confirm and refine
the earlier observation of a large North-South asymmetry in the
star counts with respect to the Galactic plane (W12). We observe a peak
asymmetry of some 10\%, that is, roughly a peak-to-trough difference 
of 20\%.   

The strong Jeans theorem states that the matter probability distribution
function in steady state 
can be written in terms of three isolating integrals. One of these 
becomes the vertical energy $E_z$, which in turn goes as $z^2$, having 
manifest even parity, as one approaches the Galactic plane \citep{binney08}. 
The appearance of our odd-parity asymmetry 
can be regarded as either falsifying the expected 
relationship  $E_z \propto z^2$
as one approaches the Galactic plane, or, alternatively, as challenging 
the notion that the matter distribution function can written  
in terms of isolating integrals at all. The ``ringing'' nature of
the asymmetry result we have found strongly favors the latter interpretation. 
We thus believe the asymmetry speaks to observational evidence
for the failure of local gravitational equilibrium, or steady-state
dynamics, in the local solar neighborhood. 

We regard this as a cautionary tale for analyses which would employ 
equilibrium assumptions to analyze the 
local distribution of dark matter.
While 20\% density differences are well within the 
uncertainties of the unknown local dark matter distribution, 
the large velocity tail 
of the local dark matter phase space distribution is likely much 
more sensitive to such effects. 
We note that recent results from the RAVE velocity 
survey show evidence for vertical ringing 
in the $z-$velocity of stars at 
the same distances discussed here \citep{williams13}.

We have considered a wide range of possible systematic effects which 
could affect the symmetry in the distribution of stars north and south of 
the Galactic plane.  We are able to quantify these systematics to a 
great extent and rule out
giant/dwarf confusion, 
photometric parallax distance 
relation differences, reddening corrections, and large scale calibration 
errors as possible sources of the observed asymmetry.  
We are able to demonstrate that photometric parallax method based
on SDSS photometry gives systematic and statistical distance errors 
of typically {\cal O}(10\%) at least for lower main sequence (K/M dwarf) populations.
To a more limited degree, we constrain the extent to which the
asymmetry can be resolved by equal-mass binary confusion, or metallicity
differences, though we cannot rule out these effects completely.  

Certain systematic effects are best explored through the study of 
distant ``blue-tip'' stars  in our sample. 
By dividing the data 
into six bins in $l$ and exploring the properties of these stars 
therein, we are able to state that there are no significant 
localized reddening correlations beyond those of the \citet{schlegel98} maps. 
We confirm the slight color difference in $(g-r)_0$
noted by \citet{schlafly10}, and we find evidence that
the most likely explanation for it is a some 0.02 mag difference in the 
global calibration of the $g$ filter in the SDSS, 
as it connects the South survey region to the 
North survey region. 
We study the asymmetries in both $(g-i)_0$ and  $(r-i)_0$ colors and show
explicitly that the asymmetry results agree nicely without regard to
which color baseline is used.
%


The $(g-r)_0,(r-i)_0$ colors of our samples are consistent with no significant 
color differences in the North versus the South. 
Regarding a possible difference in 
metallicity gradients, North and South, we do not have 
sufficient $u$-band accuracy to perform a photometric metallicity analysis 
in the manner of \citet{ivezic08}.
Slight systematic differences in the 
average metallicity of the north versus south 
populations remain possible.  However, no possible metallicity shift 
would lead to stellar density ringings confirmed here.  
We note that the persistence of the asymmetries observed in W12
in the current work, determined using a much larger sample of stars
in a slightly different portion of the sky, also argues against the
existence of idiosyncratic north-south population differences as a
source of the effect. 

In Gardner et al. (in preparation) we plan to analyze the symmetries associated
with multiple integrals of motion simultaneously, in hope of isolating
the origin of the asymmetry. 


\acknowledgements{  We acknowledge use of SDSS-III data (http://www.sdss3.org). 
 We acknowledge contributions from Kathryn Mummah on the study of
  blue tip stars vs. $l$, and we thank James Bullock for helpful comments. 
   We acknowledge the anonymous referee for several useful comments which
improved the paper.
  SG acknowledges partial support
  from the U.S. Department of Energy under contract
  DE-FG02-96ER40989 and thanks Wolfgang Korsch for expert advice
  in the use of the ROOT analysis framework. SG thanks Nora Brambilla 
  and the Excellence Cluster ``Universe'' of the Technical 
  University of Munich for hospitality during the completion of this work. 
}

\bibliographystyle{apj}

\begin{thebibliography}{34}
\expandafter\ifx\csname natexlab\endcsname\relax\def\natexlab#1{#1}\fi

\bibitem[Abazajian et al.(2009)]{abazajian09} Abazajian, K., Adelman-McCarthy, J.K., Agueros, M.A.,  et al. 2009 \apjs, 182, 543  

\bibitem[{{Aihara} {et~al.}(2011)}]{aihara11} {Aihara}, H., Allende Prieto, C., An, D. et al. 2011 \apjs, 193, 29  

\bibitem[Ahn et al.(2012)]{ahn12} Ahn, C. P., Alexandroff, R., Allende Prieto, C.  et al. 2012 \apjs, 203, 21 

\bibitem[An et al.(2007{\natexlab{a}})]{anetal07} An, D., Terndrup, D. M., Pinsonneault, M. H., Paulson, D. B., Hanson, R. B., \& Stauffer, J. R. 2007 
ApJ, 655, 233

\bibitem[An et al.(2008{\natexlab{b}})]{anmetal08} {An}, D., Johnson, J. A., Clem, J. L. et al., 2008 \apjs, 179, 326

\bibitem[Anthony-Twarog et al.(2006)]{antwarog06} 
{Anthony-Twarog}, B. J., Tanner, D., Cracraft, M., \& Twarog, B. A. 2006, \aj, 
131, 461

\bibitem[Benjamin et al.(2005)] {benetal05} Benjamin, R. A., 
Churchwell, E., Bable, B.L. et al. 2005 ApJ 630, L149

\bibitem[{{Berry} {et~al.}(2012)}]{berry12} {Berry}, M. Ivezic, Z., Sesar, B. et~al., 2012, \apj, 757, 35 

\bibitem[Betoule et al.(2013)]{betoule13} Betoule, M., Marriner, J., Regnault, N. et al. 2013 A\&A 552, 124 

\bibitem[{{Binney} \& {Tremaine}(2008)}]{binney08}
{Binney}, J., \& {Tremaine}, S. 2008, {Galactic Dynamics: Second Edition}
(Princeton University Press)


\bibitem[Bochanski et al.(2013)]{bochanski13} Bochanski, J.J., Savcheva, A., West, A.A. and Hawley, S.L. 2013 AJ 145, 40



\bibitem[Bovy \& Tremaine(2012)]{bt12} Bovy, J., \& Tremaine, S. 2012 ApJ 756, 89

\bibitem[{{Chen} {et~al.}(2011)}]{chen11} 
{Chen}, Y.~Q., Zhao, G., Carrell, K., Zhao, J.~K., 2011, \aj, 142, 8 

\bibitem[{{Covey} {et~al.}(2007)}]{covey07} 
{Covey}, K.~R., Ivezi{\'c}, Z., Schlegel, D. et al., 
2007, \aj, 134, 2398 

\bibitem[{{Dame} {et~al.}(1987)}]{dame87} {Dame}, T.~M., Ungerechts, H., Cohen, R. S. et al.  1987, \apj, 322, 706





\bibitem[Fan et al.(1996)]{fan96} Fan, X., Burstein, D., Chen., J. S. et al. 1996 \aj, 112, 628



\bibitem[Garbari et al.(2012)]{glrl12} Garbari, S., Liu, C., Read, J.I., \& Lake, G. 2012 MNRAS 425, 1445




\bibitem[{{Gilmore} {et~al.}(1989)}]{gilmore89}
Gilmore, G., Wyse, R. F. G.,  \& Kuijken, K  1989 ARA\&A, 27, 555

\bibitem[Gomez et al.(2013)]{getal13} Gomez, F.A., Minchev, I., O'Shea, B. W., Beers, T. C., Bullock, J. S., \& Purcell, C. W. 2013 MNRAS 429, 159


\bibitem[{{Humphreys} {et~al.}(2011)}]{Humphreys11} 
Humphreys, R. M., Beers, T.C., Cabanela, J.E., Grammer, S., Davidson, K., Lee, Y.S. \& Larsen, J.A. 2011 \aj, 141, 131

\bibitem[{{Ivezi{\'c}} {et~al.}(2008)}]{ivezic08} {Ivezi{\'c}}, {\v Z}., Sesar, B., Juric, M. et~al., 2008{\natexlab{a}}, \apj, 684, 287


\bibitem[{{Juri{\'c}} {et~al.}(2008)}]{juric08} {Juri{\'c}}, M., Ivezic, Z., Brooks, A. et al. 2008 \apj, 673, 864





\bibitem[Lee et al.(2008)]{lee08b} Lee, Y.~S., Beers, T.~C., 
Sivarani, T. et al.\ 2008, \aj, 136, 2050 

\bibitem[Majewski et al.(2003)] {m03} Majewski, S. R., Skrutskie, M. F., Weinberg, M. D., Ostheimer, J. C. 2003 ApJ 599, 1082

\bibitem[{{Marshall} {et~al.}(2006){Marshall}, {Robin}, {Reyl{\'e}},
  {Schultheis}, \& {Picaud}}]{marshall06}
{Marshall}, D.~J., {Robin}, A.~C., {Reyl{\'e}}, C., {Schultheis}, M., \&
  {Picaud}, S. 2006, \aap, 453, 635




\bibitem[Ostriker \& Peebles(1973)]{op73} Ostriker, J. P. and Peebles, P. J. E. 1973, ApJ 186, 467

\bibitem[Padmanabhan et al.(2008)]{Petal08} Padmanabhan, N., Schlegel, D. J., Finkbeiner, D. P. et al. 2008 ApJ 674, 1217





\bibitem[{{Raghavan} {et~al.}(2010){Raghavan}, {McAlister}, {Henry}, {Latham}, 
{Marcy}, {Mason}, {Gies}, {White}, \& {ten Brummelaar}}]{raghavan10}
{Raghavan}, D., {McAlister}, H.~A., {Henry}, T.~J., {Latham}, D.~W., 
{Marcy}, G.~W., {Mason}, B.~D., {Gies}, D.~R., {White}, R.~J., \&
{ten Brummelaar}, T.~A. 2010, \apjs, 190, 1 




\bibitem[{Schlafly} {et~al.}(2010)]{schlafly10} {Schlafly}, {Finkbeiner}, {Schlegel}, {Juri{\'c}}, {Ivezi{\'c}}, {Gibson}, {Knapp}, \& {Weaver} 2010, \apj, 725, 1175


\bibitem[Schlafly and Finkbeiner(2011)]{sf11} Schlafly, E. F., and Finkbeiner, D. P. 2011  ApJ 737, 103

\bibitem[{Schlesinger} {et~al.}(2012)]{schlesinger12} Schlesinger, K. J., Johnson, J. A., Rockosi, C. M. et al. 2012 ApJ 761, 160

\bibitem[{{Schlegel} {et~al.}(1998){Schlegel}, {Finkbeiner}, \&
  {Davis}}]{schlegel98}
{Schlegel}, D.~J., {Finkbeiner}, D.~P., \& {Davis}, M. 1998, \apj,
500, 525






\bibitem[{{Widrow} {et~al.}(2012)}]{widrow12} 
{Widrow}, L.~M., Gardner, S., Yanny, B., Dodelson, S., and Chen, H.Y., 2012, \apjl, 750, L41 (W12)


\bibitem[Williams et al.(2013)]{williams13} Williams, M.E.K., Steinmetz, M., Binney, J. et al. 2013 MNRAS, in press, arXiv:1302.2468

\bibitem[Yanny et al.(2009a)]{yanny09a} Yanny, B., Newberg, H. J., Johnson, J. A.  et al. 2009a \apj 700, 1282

\bibitem[{{Yanny} {et~al.}(2009b)}]{yanny09b} {Yanny}, B., Rockosi, C., Newberg, H. J. et~al., 2009b, \aj, 137, 4377

\end{thebibliography}

\begin{deluxetable}{lllllll}
\tablecaption{Colors of Blue-tip Stars vs. Magnitude and Longitude}
\tablehead{
\multicolumn{1}{c} {$l$ range (in ${}^\circ$)} &
\multicolumn{1}{c} {$(g-r)_0$\tablenotemark{a} N } &
\multicolumn{1}{c} {S (peak)} &
\multicolumn{1}{c} {N } &
\multicolumn{1}{c} {S } &
\multicolumn{1}{c} {N } &
\multicolumn{1}{c} {S } \\
\multicolumn{1}{c} {} &
\multicolumn{1}{c} {$[20,21]$} &
\multicolumn{1}{c} {$[20,21]$} &
\multicolumn{1}{c} {$[19,20]$} &
\multicolumn{1}{c} {$[19,20]$} &
\multicolumn{1}{c} {$[18,19]$} &
\multicolumn{1}{c} {$[18,19]$} \\
}
\startdata
60 $< l <$ 80 &      0.305 (884)&        0.316 (855)&     0.300  (673)   &   0.317 (565) & 0.291(484) & 0.306(423)\\
80 $< l <$ 100&      0.304 (733)&        0.320	(651)&   0.301 (552)      & 0.315 (482) & 0.287(400) & 0.312(342)\\
100$ < l <$ 120&     0.304 (591)&        0.333	(512)&   0.300 (465)      & 0.324 (373) &0.283(346) & 0.309(306)\\
120$ < l <$ 140&     0.305 (503)&        0.320 (481)&     0.302 (400)     &   0.325 (333) & 0.282(317) & 0.304(260)\\
140$ < l <$ 160&     { 0.306 (461)}&       {\bf 0.299 (690)}&   0.300 (367)&	0.312 (317) & 0.283(269) & 0.292(243)\\
160$ < l <$ 180&    { 0.304 (484)}&       {\bf 0.299 (784)}&     0.296 (382)&	0.306 (302) &0.294(278) & 0.294(255)\\
\enddata
\tablenotetext{a}{The peak $(g-r)_0$ color of all blue-tip (F turnoff) halo
stars are tabulated North and South of the plane in three
magnitude and six longitude bins, followed by the number of stars in 
each peak in parentheses.  The Southern tip stars
are significantly redder than the Northern tip stars in nearly all
bins, independent of magnitude and longitude -- likely indicating
a large scale calibration offset.   The numbers in boldface
indicate the direction and distance where the Southern color
is affected by Sagittarius dwarf stream stars, which are bluer
than typical halo turnoff stars.
}
\end{deluxetable}

\clearpage

\clearpage

\clearpage


\clearpage

\end{document}